\RequirePackage{lineno}
\documentclass[sort&compress,twocolumn,10pt,5p]{elsarticle}

\usepackage{afterpage}

\usepackage{atlasphysics}

\usepackage{subfigure}
\usepackage{mathrsfs}
\usepackage{graphicx}
\usepackage{multirow}
\usepackage{multirow}
\usepackage{rotating}
\usepackage{xspace}
\usepackage{slashed}
\usepackage{amsmath}

\usepackage{hyperref}

\newcommand{\alpgen}{{\tt ALPGEN}}
\newcommand{\powheg}{{\tt POWHEG}}
\newcommand{\prospino}{{\tt PROSPINO}}
\newcommand{\herwig}{{\tt HERWIG}}

\newcommand{\mcatnlo}{{\tt MC@NLO}}
\newcommand{\fewz}{{\tt FEWZ}}
\newcommand{\mcfm}{{\tt MCFM}}

\newcommand{\jimmy}{{\tt JIMMY}}
\newcommand{\sherpa}{{\tt SHERPA}}

\newlength{\myVSpace}
\newcommand{\mct}{\ensuremath{m_\mathrm{CT}}}

\def\lsim{\mathrel{\rlap{\lower4pt\hbox{\hskip1pt$\sim$}}
    \raise1pt\hbox{$<$}}}                
\def\gsim{\mathrel{\rlap{\lower4pt\hbox{\hskip1pt$\sim$}}
    \raise1pt\hbox{$>$}}}                
\makeatletter
\renewcommand{\p@subfigure}{\arabic{figure}}
\renewcommand{\thesubfigure}{\alph{subfigure}}
\renewcommand{\@thesubfigure}{(\thesubfigure)}
\makeatother

\usepackage{preprintcover}  

\PreprintCoverPaperTitle{Search for direct slepton and gaugino production in final states with two leptons and missing transverse momentum with the ATLAS detector in $pp$ collisions at $\sqrt{s}=7$~TeV}

\PreprintIdNumber{ CERN-PH-EP-2012-216}  
\PreprintCoverAbstract{
A search for the electroweak pair production of charged sleptons and weak gauginos decaying into final 
states with two leptons is performed using 4.7~fb$^{-1}$ of proton-proton collision data at $\sqrt{s}=7$~TeV recorded with
the ATLAS experiment at the Large Hadron Collider.  No significant excesses are observed with respect to the prediction from Standard
Model processes. In the scenario of direct slepton production, if the sleptons decay directly into the lightest neutralino, 
left-handed slepton masses between 85 and 195~GeV are excluded at 95\% confidence level for a 20~GeV neutralino. Chargino masses between 110 and 340~GeV are 
excluded in the scenario of direct production of wino-like chargino pairs decaying into the lightest neutralino via an intermediate on-shell 
charged slepton for a 10~GeV neutralino. The results are also interpreted in the framework of the phenomenological minimal supersymmetric Standard Model.
}\PreprintJournalName{Physics Letters B}  

\begin{document}

\begin{frontmatter}

\title{\vspace{-0.5cm} \flushright{\normalsize{CERN-PH-EP-2012-216}} \\ \center{Search for direct slepton and gaugino production in final states with two leptons and missing transverse momentum with the ATLAS detector in $pp$ collisions at $\sqrt{s}=7$~TeV}}
\journal{Physics Letters B}

\author{The ATLAS Collaboration}

\begin{abstract}
A search for the electroweak pair production of charged sleptons and weak gauginos decaying into final 
states with two leptons is performed using 4.7~fb$^{-1}$ of proton-proton collision data at $\sqrt{s}=7$~TeV recorded with
the ATLAS experiment at the Large Hadron Collider.  No significant excesses are observed with respect to the prediction from 
Standard Model processes. In the scenario of direct slepton production, if the sleptons decay directly into the lightest neutralino, 
left-handed slepton masses between 85 and 195~GeV are excluded at 95\% confidence level for a 20~GeV neutralino. Chargino masses 
between 110 and 340~GeV are excluded in the scenario of direct production of wino-like chargino pairs decaying into the lightest 
neutralino via an intermediate on-shell charged slepton for a 10~GeV neutralino. The results are also interpreted in the framework 
of the phenomenological minimal supersymmetric Standard Model.
\end{abstract}

\end{frontmatter}

\section{Introduction}

Weak scale Supersymmetry (SUSY)~\cite{Miyazawa:1966,Ramond:1971gb,Golfand:1971iw,Neveu:1971rx,Neveu:1971iv,Gervais:1971ji,Volkov:1973ix,Wess:1973kz,Wess:1974tw} 
is an extension to the Standard Model (SM). It postulates for each known boson or 
fermion the existence of a particle whose spin differs by one-half unit from the SM 
partner. The introduction of these new particles provides solutions to the hierarchy problem~\cite{Weinberg:1975gm,Gildener:1976ai,Weinberg:1979bn,Susskind:1978ms}
and, if R-parity is conserved~\cite{Fayet:1976et,Fayet:1977yc,Farrar:1978xj,Fayet:1979sa,Dimopoulos:1981zb}, 
a dark matter candidate in the form of the lightest supersymmetric particle (LSP). 
R-parity conservation is assumed in this letter, hence SUSY particles are always produced in pairs. 
In a large fraction of the SUSY parameter space the LSP is the weakly interacting lightest neutralino, $\tilde{\chi}^{0}_{1}$.

Gluinos ($\tilde{g}$) and squarks ($\tilde{q}$) are the SUSY partners of gluons and quarks.
Charginos ($\tilde{\chi}^\pm_i$, $i=1,2$) and neutralinos ($\tilde{\chi}^0_j$, $j=1,2,3,4$) are the mass 
eigenstates formed from the linear superposition of the SUSY partners of the Higgses and electroweak gauge bosons: 
higgsinos, winos and the bino (collectively, gauginos). The SUSY partners of the charged leptons are the selectron,
smuon and stau, collectively referred to as charged sleptons ($\tilde{\ell}^\pm$). The SUSY partners of the 
standard model left-handed leptons are referred to as left-handed sleptons. If the masses of the gluinos and squarks are greater than a few~TeV and
the weak gauginos and sleptons have masses of a few hundreds of GeV, the direct production of weak gauginos 
and sleptons may dominate the production of SUSY particles at the Large Hadron Collider (LHC). Such a scenario is 
possible in the general framework of the phenomenological minimal supersymmetric SM (pMSSM)~\cite{pMSSM}. 
Naturalness suggests that third generation sparticles, charginos and 
neutralinos should have masses of a few hundreds of GeV~\cite{Barbieri:1987fn,deCarlos:1993yy}. Light sleptons are expected in 
gauge mediated~\cite{gmsb1} and anomaly mediated~\cite{amsb1,amsb2} SUSY breaking scenarios. 
Light sleptons could also play a role in helping SUSY 
to provide a relic dark matter density consistent with observations~\cite{slepton_dm1,slepton_dm2}.

This letter presents the first search for direct left-handed slepton pair production at the LHC,
and a dedicated search for direct chargino pair production in final states with two leptons 
(electrons, $e$, or muons, $\mu$). Searches for the general pair production of gauginos decaying into 
two-lepton final states are also presented. The analysis presented in this letter is not sensitive to right-handed slepton pair production which has much lower cross-section.

\subsection{Direct Slepton and Chargino Pair Production}
Sleptons can be produced directly in a process similar to Drell-Yan production~\cite{sleptons}. 
The search in this letter targets the direct pair production of left-handed charged sleptons, where 
each charged slepton $\tilde{\ell}$ (selectron or smuon) decays through $\tilde{\ell}^{\pm} \to {\ell}^\pm \tilde{\chi}^0_1$, 
yielding a final state with two same flavour (SF) charged leptons. The undetected $\tilde{\chi}^0_{1}$ 
gives rise to large missing transverse momentum in the event.
Previous experimental searches for direct slepton production~\cite{lepsusy} assumed gaugino unification. 
In the present work this assumption is dropped, thereby removing the lower limit on the mass of the ${\tilde{\chi}^0_1}$.
Direct chargino pair production, where each chargino decays through 
$\tilde{\chi}^{\pm}_{1}\rightarrow {\ell}^\pm\nu\tilde{\chi}^{0}_{1}$ leads to a signature similar
to that of slepton pair production. The analysis presented also targets this production channel and subsequent decay, 
setting limits on the chargino mass, without the assumptions on the mass of the $\tilde{\chi}^0_2$ 
usually present in trilepton searches.

\subsection{Other Weak Gaugino Production}
In the general framework of the pMSSM, several weak gaugino production channels can lead 
to final states with two leptons. Production modes such as
$\tilde{\chi}^{0}_2\tilde{\chi}^\pm_{i=1,2}$ or $\tilde{\chi}^{0}_2\tilde{\chi}^0_{j=2,3,4}$, 
with the subsequent decays $\tilde{\chi}^{0}_{2} \rightarrow {\ell}^\pm {\ell}^\mp \tilde{\chi}^{0}_{1}$
and $\tilde{\chi}^{0}_{j},\tilde{\chi}^{\pm}_{i} \rightarrow q \bar{q'} \tilde{\chi}^{0}_{1}$ 
are addressed by a signal region containing two leptons and two jets. In order to complement 
existing and future trilepton searches a dedicated signal region with two same charge 
leptons is designed to be sensitive to trilepton final states from
$\tilde{\chi}^0_2\tilde{\chi}^\pm_1\rightarrow({\ell}^\mp {\ell}^\mp\tilde{\chi}^0_1)+({\ell}^\mp \nu\tilde{\chi}^0_1)$ 
where one lepton is not identified. All final states yield missing transverse energy
due to the presence of two $\tilde{\chi}^0_{1}$'s.

Model-independent visible cross-section upper limits are obtained in each signal region to
address the large variety of possible production and decay modes in the gaugino sector. 
The results are also interpreted in the framework of the pMSSM.
This search is not sensitive to weak gaugino decays via on-shell $Z$~bosons. Previous limits 
on weak chargino and neutralino production have been placed at LEP~\cite{lepsusy}, the Tevatron~\cite{cdftrilepton,D0trilepton} 
and at the LHC~\cite{2lep2011,3lep2011}.

\section{The ATLAS Detector}

The ATLAS experiment~\cite{Aad:2008zzm} is a multi-purpose particle
physics detector with a forward-backward symmetric cylindrical
geometry and nearly 4$\pi$ coverage in solid angle.\footnote{
ATLAS uses a right-handed coordinate system with its origin at the nominal
interaction point in the centre of the detector and the $z$-axis along the
beam pipe. Cylindrical coordinates $(r,\phi)$ are used in the transverse
plane, $\phi$ being the azimuthal angle around the beam pipe. The pseudorapidity $\eta$ is
defined in terms of the polar angle $\theta$ by $\eta=-\ln\tan(\theta/2)$.
}  It contains four superconducting magnet systems, which include a 
thin solenoid surrounding the inner tracking detector (ID), and barrel and end-cap toroids supporting a muon spectrometer. The ID occupies the pseudorapidity region  $\left|\eta\right| <2.5$ and consists of a silicon pixel detector, a silicon microstrip detector (SCT), and a transition
radiation tracker (TRT). In the pseudorapidity
region $\left|\eta\right| < 3.2$, high-granularity liquid-argon (LAr)
electromagnetic (EM) sampling calorimeters are used.  An iron-scintillator tile
calorimeter provides coverage for hadron detection over
$\left|\eta\right| < 1.7$.  The end-cap and forward regions, spanning
$1.5~<~\left|\eta\right|~<~4.9$, are instrumented with LAr calorimeters
for both EM and hadronic measurements. The muon spectrometer surrounds the calorimeters and consists of a system of precision tracking chambers ($|\eta|<2.7$), and detectors for triggering ($|\eta|<2.4$).

\section{Simulated Samples}

\label{sec:MC}

\subsection{Standard Model Production}

Monte Carlo (MC) simulated event samples are used to develop and validate the analysis procedure 
and to evaluate the SM backgrounds in the signal region. The dominant backgrounds include fully-leptonic 
$t\bar{t}$, $Z/\gamma^*+$jets, single top and dibosons ($WW$, $WZ$ and $ZZ$). 
Production of top quark pairs is simulated with \powheg~\cite{powheg}, using a top quark mass of $172.5$~GeV. 
Samples of $W$ to $l\nu$ and $Z/\gamma^*$ to $ll$, produced with accompanying jets (of both light and heavy 
flavour), are obtained with \alpgen~\cite{Mangano:2002ea}. Diboson ($WW$, $WZ$, $ZZ$) production is simulated
with \sherpa~\cite{sherpa} in signal regions requiring jets and with \herwig~\cite{herwig} elsewhere. Single 
top production is modelled with \mcatnlo~\cite{mcatnlo1,mcatnlo2,mcatnlo3}. Fragmentation and hadronisation for the \alpgen~and \mcatnlo~samples 
are performed with \herwig, using \jimmy~\cite{Butterworth:1996zw} for the underlying event, and with {\tt PYTHIA}~\cite{pythia} 
for the~\powheg~sample. Expected diboson yields are normalised using NLO QCD predictions obtained with \mcfm~\cite{Campbell:1999ah,Campbell:2011bn}. The top-quark contribution is normalised to approximate next-to-next-to-leading order (NNLO) calculations~\cite{Aliev:2010zk}. The inclusive $W$ and $Z/\gamma^*$ production cross-sections are normalised to the next-to-next-to-leading order (NNLO) cross-sections obtained using \fewz~\cite{Gavin:2010az}. \mcatnlo~samples are used to assess the systematic uncertainties associated with the choice of generator for $t\bar{t}$ production, and {\tt AcerMC}~\cite{Kersevan:2004yg} samples are used to assess the uncertainties associated with initial and final state radiation (ISR/FSR)~\cite{ATLAS:2012al}. \alpgen,~\herwig~ and \sherpa~samples are used to assess the systematic uncertainties associated with the choice of generator for diboson production. \sherpa~ is used to evaluate the small contribution from internal conversions.

\subsection{Direct Slepton and Direct Gaugino Production}

Four signal regions are designed, optimised for the discovery of various SUSY 
models where sleptons or gauginos are directly produced in the $pp$ interaction. The pMSSM 
framework is used to produce two sets of signal samples, one where sleptons are directly produced and
one where gauginos are directly produced. These samples are used to set the limits on the masses
of the directly produced sleptons and gauginos. Samples are also produced in a simplified model 
at given LSP and chargino masses, and are then used to set limits on the chargino mass, 
independently of the $\tilde{\chi}^0_2$ mass. In all SUSY models the masses of the squarks,
gluinos and third generation supersymmetric partners of the fermions are large (2.5 TeV in the 
direct slepton production pMSSM models and 2~TeV in the direct gaugino pMSSM and simplified models).

The direct slepton models are based on those described in Ref.~\cite{benchmark_for_lhc}.
Masses of all gauginos apart from the $\tilde{\chi}^0_1$ are set to 2.5~TeV. The sensitivity of the present search 
to a given model is determined by the slepton production cross-section and by the mass of the $\tilde{\chi}^0_1$,
which affects the kinematics of the final state leptons. The mass of the  bino-like $\tilde{\chi}^0_1$  is 
varied by scanning values of gaugino mass parameter $M_1$ in steps of 20~GeV in the range 20--160~GeV.
The common selectron and smuon mass is generated in the range 70--190~GeV, scanned in steps of 
20~GeV with the constraint $m_{\tilde{\ell}}>$ $m_{\tilde{\chi}^0_1}+30$~GeV. The cross-section for direct slepton pair production in these models decreases from 3.9 to 0.05~pb independently of neutralino mass as the slepton mass increases from 70 to~190~GeV.

In the considered simplified models, the masses of the relevant particles ($\tilde{\chi}^0_1$, $\tilde{\nu}$, $\tilde{\ell}_L$, $\tilde{\chi}^\pm_1$ and $\tilde{\chi}^0_2$) are the only free parameters.
The latter are wino-like and
$\tilde{\chi}^0_1$ is bino-like. The $\tilde{\chi}^\pm_1$ are pair-produced  
via the $s$-channel exchange of a virtual gauge boson and decay via left-handed sleptons, 
including $\tilde{\tau}$, and $\tilde{\nu}$ of mass  
$m_{\tilde{\nu}}=m_{\tilde{\ell}_L}\,$$=\,$$(m_{\tilde{\chi}_1^{0}} + m_{\tilde{\chi}_1^{\pm}} )/2$ 
with a branching ratio of 50\%  each. 
The cross-section for $\tilde{\chi}^\pm_1\tilde{\chi}^\mp_1$ pair production in these models is as high as 3~pb
for a chargino mass of 50~GeV and decreases rapidly at higher masses, reaching below $\sim0.2$~pb for masses above 200~GeV.

For the other weak gaugino production channels, a set of pMSSM models with intermediate sleptons 
in the gaugino decay chain are generated. The right-handed sleptons, with a common mass for 
all three generations, are inserted halfway between the two lightest neutralino masses 
while left-handed slepton masses are kept beyond reach. 

The masses of the charginos and the neutralinos depend on the gaugino and Higgsino mass parameters $M_1$, $M_2$ and $\mu$ and the ratio of the expectation values of the two Higgs doublets ($\tan\beta$) via mixing matrices~\cite{Haber:1984rc}. The chargino masses are given by the solution of a $2\times2$ matrix equation which is dependent on $M_2$, $\mu$ and $\tan\beta$~\cite{Beringer:1900zz}. The neutralino masses are found by solving a $4\times4$ matrix equation; solutions to which are given in Refs.~\cite{Choi:2001ww,ElKheishen:1992yv,Guchait:1991ia}. The parameters $M_1$, $M_2$ and $\mu$ are varied independently while $\tan\beta$ is set to 6.
In the pMSSM model the cross-sections vary significantly 
(between 0.5 and 100~pb for $M_1=250$~GeV, with the highest cross-sections at low $M_2$ and $\mu$). 
The present direct gaugino production search is only sensitive to models with intermediate sleptons. 

Signal samples for the pMSSM and slepton model points are generated with {\tt HERWIG}, whereas {\tt Herwig++}~\cite{Bahr:2008pv} is used to generate the simplified model points.
Signal cross-sections are calculated to next-to-leading order in the strong coupling constant (NLO) using \prospino~\cite{Beenakker:1996ch}. The nominal cross-section and the uncertainty are taken from an envelope of cross-section predictions using different parton distribution function (PDF) sets and factorisation and renormalisation scales, as described in Ref.~\cite{Kramer:2012bx}. 

All MC samples are produced using a {\tt GEANT4}~\cite{geant4} based detector simulation~\cite{atlassimulation}. 
The effect of multiple proton-proton collisions from the same or different bunch crossings is incorporated into the simulation by overlaying additional minimum bias events onto hard scatter events using {\tt PYTHIA}. Simulated events are weighted to match the distribution of the number of interactions per bunch crossing observed in data.

\section{Data and Event Selection}

\label{sec:objectreconstruction} 

The 7~TeV proton-proton collision data analysed were recorded between March and October 2011. Application of beam, detector and data-quality requirements yields a total integrated luminosity of 4.7~fb$^{-1}$. Events are triggered using a combination of single and double lepton triggers. The single electron triggers vary with the data taking period, and the tightest trigger has an efficiency of $\sim$97\% for offline electrons with $p_\mathrm{T}>25$~GeV.  The single muon trigger used for all data taking periods reaches an efficiency plateau of $\sim$75\% ($\sim$90\%) in the barrel (end-caps) for muons with $p_\mathrm{T}>20$~GeV. All quoted efficiencies have been measured with respect to reconstructed leptons.
The double lepton triggers reach similar plateau efficiencies, but at lower $p_\mathrm{T}$ thresholds: $>17$~GeV for the dielectron trigger, and $>12$~GeV for the dimuon trigger; for the electron-muon trigger the thresholds are 15 and 10~GeV respectively. 
One or two signal leptons are required to have triggered the event, and be matched to the online triggered leptons: one lepton if one is above the appropriate single lepton trigger plateau threshold, or two leptons if there is no such lepton.
An exception to this rule is applied in the $\mu\mu$ channel. In this case when one lepton is above the single lepton trigger plateau threshold, and the other above the double lepton threshold, a logical OR of both triggers is used to recover efficiency.

Jet candidates are reconstructed using the anti-$k_t$ jet clustering algorithm~\cite{Cacciari:2008gp} with a distance parameter of $0.4$. 
The jet candidates are corrected for the effects of calorimeter non-compensation and inhomogeneities by using $p_\mathrm{T}$ and $\eta$-dependent calibration factors based on MC simulations and validated with extensive test-beam and collision-data studies \cite{Aad:2011he}. 
Only jet candidates with transverse momenta $p_\mathrm{T} > 20$ \GeV{} and $|\eta|<4.5$ are subsequently retained. Jets likely to have arisen from detector noise or cosmic rays are rejected~\cite{Aad:2011he}. Electron candidates are required to have $p_\mathrm{T} > 10~\GeV$, $|\eta| < 2.47$, and pass the ``medium''  shower shape and track selection criteria of Ref.~\cite{Aad:2010yt}. Muon candidates are reconstructed using either a full muon spectrometer track matched to an ID track, or a partial muon spectrometer track matched to an ID track. They are then required to have $p_\mathrm{T} > 10$ \GeV{} and $|\eta| < 2.4$. They must be reconstructed with sufficient hits in the pixel, SCT and TRT detectors. 

The measurement of the missing transverse momentum two-vector, $\mathbf{p}_\mathrm{T}^\mathrm{miss}$, and its magnitude, $E_\mathrm{T}^\mathrm{miss}$, is based on the transverse momenta of all electron and muon candidates, all jets, and all clusters of calorimeter energy with $|\eta|<4.9$ not associated to such objects. The quantity $E^\mathrm{miss,rel.}_\mathrm{T}$~is defined as:
\begin{equation}
	E^\mathrm{miss,rel.}_\mathrm{T} =  \left\{   
	\begin{array}{l l}  
	E^\mathrm{miss}_\mathrm{T}                                      & \quad \text{if $\Delta \phi_{\ell,j} \geq \pi$/2} \\
	E^\mathrm{miss}_\mathrm{T} \times \sin{\Delta \phi_{\ell,j}} & \quad \text{if  $\Delta \phi_{\ell,j} < \pi$/2} \\
       \end{array} \right., \nonumber
\end{equation}
where $\Delta \phi_{\ell,j}$ is the azimuthal angle between the direction of 
$\textbf{p}^\mathrm{miss}_\mathrm{T}$ and that of the nearest electron, muon or jet. 
In a situation where the momentum of one of the jets or leptons is significantly mis-measured, 
such that it is aligned with the direction of $\textbf{p}^\mathrm{miss}_\mathrm{T}$, 
only the \met ~component perpendicular to that object is considered. This is used to significantly 
reduce mis-measured $E^\mathrm{miss}_\mathrm{T}$ in processes such as 
$Z/\gamma^*\rightarrow e^{+} e^{-}, \mu^{+} \mu^{-}$~\cite{Aad:2011kk}. 

Signal electrons, muons and jets are then selected. Signal electrons are further required to pass the ``tight''~\cite{Aad:2010yt} quality criteria, which place additional requirements on the ratio of calorimetric energy to track momentum, and the number of high-threshold hits in the TRT. They are also required to be isolated: the $p_\mathrm{T}$ sum of tracks above 1~GeV within a cone of size $\mathrm{\Delta} R = \sqrt{(\mathrm{\Delta}\eta)^2+(\mathrm{\Delta}\phi)^2}=0.2$ around each electron candidate (excluding the electron candidate itself) is required to be less than 10\% of the electron $p_\mathrm{T}$. Signal muons must also be isolated: the $p_\mathrm{T}$ sum of tracks within a cone of size $\mathrm{\Delta} R=0.2$ around the muon candidate is required to be less than 1.8~GeV.

Signal jets are subject to the further requirements $p_\mathrm{T}>30$~GeV, $|\eta|<2.5$ and 
a ``jet vertex fraction'' greater than 0.75. The jet vertex fraction is defined as the total 
track momentum associated to the jet and coming from the primary vertex divided by the total track 
transverse momentum in the jet.

The jet vertex fraction quantifies the fraction of track transverse momentum from the primary vertex, associated to a jet. This variable is used to remove jets that originated from other collisions, and also discards jets without reconstructed tracks.

A $b$-tagging algorithm~\cite{ATLAS-CONF-2011-102}, which exploits the long lifetime of weak $b$- and $c$-hadron decays inside a candidate jet, is used to identify jets containing a $b$-hadron decay. The mean nominal $b$-tagging efficiency, determined from $t\bar{t}$ MC events, is 80\%, with a misidentification (mis-tag) rate for light-quark/gluon jets of less than 1\%. Scale factors (which depend on $p_\mathrm{T}$ and $\eta$) are applied to all MC samples to correct for small discrepancies in the $b$-tagging performance observed in data with respect to simulation.
 
Basic data quality requirements are then applied. Selected events in each signal region (SR-) and control region must satisfy the following requirements. The primary vertex in the event must have at least five associated tracks and each event must contain exactly two signal leptons of opposite sign (OS) or same sign (SS). Both of these leptons must additionally satisfy the full list of lepton requirements, and the dilepton invariant mass, $m_{\ell\ell}$, must be greater than $20$~GeV across all flavour combinations. 

\section{Signal Regions}

\begin{table}[!tp]
\begin{center}
\begin{tabular}{|c|c|}
\hline
Targeted Process & Signal Region\\
\hline
\hline
\multicolumn{2}{|c|}{Two-lepton Final States} \\
\hline
$\tilde{\ell}^\pm \tilde{\ell}^\mp \rightarrow ({\ell}^\mp \tilde{\chi}^0_1) + ({\ell}^\mp \tilde{\chi}^0_1)$               & SR-$m_\textup{T2}$ \\
\hline
$\tilde{\chi}^\pm_1\tilde{\chi}^\mp_1\rightarrow({\ell}^\mp\nu\tilde{\chi}^0_1)+({\ell}^\mp\nu\tilde{\chi}^0_1)$      & SR-$m_\textup{T2}$, SR-OSjveto \\
\hline
$\tilde{\chi}^{0}_2\tilde{\chi}^{}_{i} \rightarrow ({\ell}^\mp {\ell}^\mp\tilde{\chi}^0_1)+(q\bar{q}^{'}\tilde{\chi}^0_1)$ & SR-2jets \\
\hline
\multicolumn{2}{|c|}{Three-lepton Final States} \\
\hline
$\tilde{\chi}^0_2\tilde{\chi}^\pm_1\rightarrow({\ell}^\mp {\ell}^\mp\tilde{\chi}^0_1)+({\ell}^\mp \nu\tilde{\chi}^0_1)$ & SR-OSjveto, SR-SSjveto \\
\hline
\end{tabular}
\caption{Decay modes targeted by each signal region, $\tilde{\chi}^{}_{i}$ denotes either a chargino or a neutralino. 
In decays producing three real leptons, one must be mis-reconstructed or fall outside the acceptance of the detector.  \label{SR_targets}}
\end{center}
\end{table}

In this analysis four signal regions are defined.
The first and main signal region (labelled SR-$m_\textup{T2}$) exploits the ``stransverse'' mass variable, $m_\textup{T2}$~\cite{Barr:2003rg,Lester:1999tx}, to provide sensitivity to both $\tilde{\chi}^\pm_1$ and 
$\tilde{\ell}^\pm$ pair production. This variable is defined as:
$m_\textup{T2}=\min\limits_{{\textbf{q}}_\textup{T}+{\textbf{r}}_\textup{T}=\textbf{p}^\textup{miss}_\textup{T}}
\left[\textup{max}\left(m_{T}({\textbf{p}}^{\ell_1}_\textup{T},\textbf{q}_\textup{T}),m_{T}({\textbf{p}}^{\ell_2}_\textup{T},\textbf{r}_\textup{T})\right)\right],$
\label{eq:mt2}

where ${\textbf{p}}^{\ell_1}_\textup{T}$ and $\textbf{p}^{\ell_2}_\textup{T}$ are the transverse momenta of the two leptons, and $\textbf{q}_\textup{T}$ and $\textbf{r}_\textup{T}$ are two vectors which satisfy $\textbf{q}_\textup{T}+{\textbf{r}}_\textup{T}=\textbf{p}^\textup{miss}_\textup{T}$. $m_\textup{T}$ indicates the transverse mass, $m_\textup{T}=\sqrt{2E^{\ell_1}_{\textup{T,p}}E_{\textup{T,q}}(1-\cos\phi)}$, where $E_\textup{T}$ is the transverse energy of a particle and $\phi$ the angle between the two particles in the transverse plane. The minimisation is performed over all possible decompositions of $\textbf{p}^\textup{miss}_\textup{T}$.

The search for $\tilde{\ell}^\pm$  pair production uses only the same flavour
channels $e^{+}e^{-}$ and $\mu^{+}\mu^{-}$, while the $\tilde{\chi}^\pm_1$ pair production search
also relies on $e^{\pm}\mu^{\mp}$. Additional sensitivity to $\tilde{\chi}^\pm_1$ pair production is 
provided by the second signal region, SR-OSjveto, which selects OS lepton pairs with high $E^\mathrm{miss}_\mathrm{T}$ in events with no signal jets. 

The production modes $\tilde{\chi}^{0}_2\tilde{\chi}^\pm_{i}$ or $\tilde{\chi}^{0}_2\tilde{\chi}^0_i$,
with the subsequent decays $\tilde{\chi}^{0}_{2} \rightarrow {\ell}^\pm {\ell}^\mp \tilde{\chi}^{0}_{1}$
and $\tilde{\chi}^{0}_{i},\tilde{\chi}^{\pm}_{i} \rightarrow q \bar{q'} \tilde{\chi}^{0}_{1}$
are targeted by a region called SR-2jets, which selects events with two signal jets and two OS leptons.

In this letter the region SR-OSjveto and an equivalent region, SR-SSjveto, which instead selects the events with SS lepton pairs, also target a {three} 
lepton final state. The explicit veto in this analysis on a third lepton makes the results 
in these regions orthogonal to results from direct gaugino searches with three or more leptons~\cite{3lep2011}. These regions recover events which are not reconstructed in searches for $\geq3$ leptons because one of the three leptons falls outside the acceptance of the detector and selection criteria. The processes directly targeted by each signal region are stated explicitly in Table~\ref{SR_targets}.

The exact requirements on the values to be taken by each variable in each signal region were determined by optimising the 
expected reach using a significance measure~\cite{CSC} in either the neutralino--slepton mass plane of the pMSSM model (SR-$m_\textup{T2}$), the neutralino--chargino 
mass plane of the simplified model (SR-OSjveto and SR-SSjveto) or the $M_1$--$\mu$ mass plane of the pMSSM (SR-2jets). 
Table~\ref{SR} summarises the requirements for entering each signal region. 

\begin{table}[!tp]
\begin{center}
\begin{tabular}{|c|c|cc|c|}
\hline
SR-                               &$m_\mathrm{T2}$ & OSjveto      & SSjveto &2jets  \\
 \hline
 \hline 
 charge                            & OS                 & OS       &  SS                       & OS      \\ 
 flavour                           & any                & \multicolumn{2}{c|}{any}         & SF      \\ 
 $m_{\ell\ell}$                          & $Z$-veto           & $Z$-veto &      -              & $Z$-veto \\ 
 signal jets                       & $=0$               & \multicolumn{2}{c|}{$=0$}             & $\geq2$ \\ 
 signal $b$-jets                   & -                  & \multicolumn{2}{c|}{-} & $=0$    \\ 
 $E^\mathrm{miss,rel.}_\mathrm{T}$ & $>40$              & \multicolumn{2}{c|}{$>100$}        & $>50$  \\ 
 other                             & $m_\textup{T2}>90$ & \multicolumn{2}{c|}{-}          & $m_\textup{CT}$-veto \\
 \hline
\end{tabular}
\caption{\label{SR}Signal regions. OS (SS) denotes two opposite-sign (same-sign) signal leptons, of same (SF) or different (DF) flavour. The $Z$-veto rejects events with $m_{\ell\ell}$ within 10~GeV of the $Z$ boson mass (91.2~GeV). The $m_\textup{CT}$-veto rejects events kinematically consistent with $t\bar{t}$ (Section~\ref{mct}). The values quoted for $E^\mathrm{miss,rel}_\mathrm{T}$ and $m_\textup{T2}$ are in units of~GeV.}
\end{center}
\end{table}

\subsection{Direct Slepton and Chargino Pair Production}

In SR-$m_\textup{T2}$ the properties of $m_\textup{T2}$ are exploited to search for 
$\tilde{\ell}^\pm\tilde{\ell}^\mp$ and $\tilde{\chi}^\pm_1\tilde{\chi}^\mp_1$ production followed 
by decay to final states containing exactly two OS leptons (of different flavour, DF, or same 
flavour, SF), no signal jets, and $E^\mathrm{miss}_\mathrm{T}$ from the two $\tilde{\chi}^0_1$. 
In this signal region $t\bar{t}$ and $WW$ are dominant backgrounds. For large mass differences between the 
sleptons (charginos) and the lightest neutralino, the $m_\textup{T2}$ distribution for signal 
events extends significantly beyond the distributions for $t\bar{t}$ and diboson backgrounds. 
The optimised value for the lower $m_\textup{T2}$ requirement is 90~GeV, just above the $W$ boson mass 
(which is the approximate end-point of the $WW$ and $t\bar{t}$ distributions). 

A rejection of events with $m_{\ell\ell}$ within 10~GeV of the $Z$~mass reduces $Z/\gamma^*$ backgrounds. For the direct slepton pMSSM models with a 20~GeV neutralino, the product of the kinematic and geometrical acceptance and reconstruction and event selection efficiencies varies between 0.1 and 4.0\% in this SR for slepton masses between 90 and 190~GeV. For fixed 190~GeV slepton mass, this product increases from 0.2 to 4.0\% as the neutralino mass decreases from 140 to 20~GeV. In the simplified models, for $\tilde{\chi}^\pm_1\tilde{\chi}^\mp_1$ pair production, the product of acceptance and efficiency ranges between 1 and 7\%, increasing towards higher chargino and lower neutralino masses.

In SR-OSjveto a different approach to reducing the backgrounds is taken. The $m_\textup{T2}$ variable is not used, and instead more stringent requirements are replaced on $E^\mathrm{miss,rel.}_\mathrm{T}$ to suppress the $t\bar{t}$ background.
The dominant $Z$ background is suppressed by rejecting events with $m_{\ell\ell}$ within 10~GeV of the $Z$ boson mass.
The final requirement, on $E^\mathrm{miss,rel.}_\mathrm{T}$, further increases sensitivity to the signals which are associated with much higher $E^\mathrm{miss}_\mathrm{T}$ than the SM backgrounds. In the simplified models, for $\tilde{\chi}^\pm_1\tilde{\chi}^\mp_1$ pair production, the product of acceptance and efficiency ranges between 1 and 8\%, increasing towards higher chargino and lower neutralino masses.

In SR-$m_\textup{T2}$ the expected number of direct slepton signal events for 
$m_{\tilde{\ell}}=130$~GeV and $m_{\tilde{\chi}^{0}_{1}}=20$~GeV is 20.7$\pm$0.8(syst)$\pm$0.6(theory), where the
first uncertainty denotes experimental uncertainties detailed below, while the theory uncertainty contains
PDF and scale uncertainties. In SR-OSjveto the expected number of direct 
chargino pair events with $m_{\tilde{\chi}^{\pm}_{1}}=175$~GeV 
and $m_{\tilde{\chi}^{0}_{1}}=25$~GeV is 67.8$\pm$3.4(syst)$\pm$2.3(theory).

\subsection{Other Weak Gaugino Production}
\label{mct}

In the production channel and decay 
$\tilde{\chi}^{0}_2\tilde{\chi}^{}_{i} \rightarrow ({\ell}^\mp {\ell}^\pm\tilde{\chi}^0_1)+(q\bar{q}^{'}\tilde{\chi}^0_1)$
the resulting OS two lepton final state has significant $E^\mathrm{miss}_\mathrm{T}$ and at least two signal jets. The region SR-2jets is thus sensitive to these decays. In SR-2jets, top background is reduced using a ``top-tag'' veto. The top-tagging requirement is imposed through the use of the contransverse mass variable $\mct$ \cite{Tovey:2008}. This observable can be calculated from the four-momenta of the selected signal jets and leptons:
\begin{align}
  \mct^{2}(v_{1},v_{2}) = \left[ E_{\rm T}(v_{1}) + E_{\rm T}(v_{2}) \right]^2 \notag\\
  -  \left[  {\bf p_{\rm T}}(v_{1}) - {\bf p_{\rm T}}(v_{2}) \right]^2 , \nonumber
  \label{eq:mct_endpoints}
\end{align}
where $v_i$ can be a lepton ($l$), jet ($j$) or a lepton-jet combination. Transverse momentum vectors 
are defined by ${\bf p_{\rm T}}$ and transverse energies $E_\mathrm{T}$ are defined as $E_\mathrm{T} = \sqrt{p_{\rm T}^2 +m^2}$. 
The quantities $m_{\textup{CT}}(j,j)$, $m_{\textup{CT}}(l,l)$ and $m_{\textup{CT}}(jl,jl)$ are bounded from 
above by analytical functions of the top quark and $W$ boson masses. A top-tagged event must have at least 
two jets with $p_\mathrm{T}>30$~GeV, and the scalar sum of the $p_\mathrm{T}$ of at least one combination 
of two signal jets and the two signal leptons in the event must exceed 100~GeV. Furthermore, top-tagged 
events are required to possess $\mct$ values calculated from combinations of signal jets and leptons consistent 
with the expected bounds from $t\bar{t}$ events as described in Ref.~\cite{Polesello:2009rn}. 
Further top rejection is achieved using a veto on any events containing a signal jet tagged as a $b$-jet. $Z$ backgrounds are reduced using the $Z$-veto, and 
sensitivity increased by searching at high-$E^\mathrm{miss,rel.}_\mathrm{T}$.
The expected number of signal 
events in SR-2jets from 
$\tilde{\chi}^{0}_2\tilde{\chi}^{\pm}_{1} \rightarrow ({\ell}^\mp {\ell}^\pm\tilde{\chi}^0_1)+(q\bar{q}^{'}\tilde{\chi}^0_1)$ 
for a pMSSM point with $M_1=100$~GeV, $M_2=120$~GeV, $\mu=100$~GeV is 37.6$\pm$4.9(syst)$\pm$7.0(theory).

In the regions targeting fully-leptonic $\tilde{\chi}^0_2\tilde{\chi}^\pm_1$ decays (SR-OSjveto and SR-SSjveto), a veto on events containing a signal jet reduces hadronic backgrounds, and high $E^\mathrm{miss,rel.}_\mathrm{T}$ increases the sensitivity to SUSY decays. The final state leptons can be of either OS or SS. In the absence of significant expected $Z$ background in the SS signal region, no $Z$-veto is applied.
The expected number of signal events in SR-SSjveto from fully-leptonic $\tilde{\chi}^0_2\tilde{\chi}^\pm_1$ decays
for a simplified model with $m_{\tilde{\chi}^0_2}=m_{\tilde{\chi}^\pm_1}=200$~GeV and $m_{\tilde{\chi}^0_1}=50$~GeV,
is  12.4$\pm$1.4(syst)$\pm$0.7(theory).

\section{Background Evaluation}\label{sec:backgrounds}

\begin{table}[!tp]
\begin{center}
\begin{tabular}{|c|c|c|c|c|}
\hline
 & {top} & $WW$ & {$Z+X$ } \\
 \hline 
 \hline
 $m_{\ell\ell}$                                & $Z$-veto                   & $Z$-veto                      & $Z$-window          \\
 signal jets                             & $\geq2$                    & =0                            & $=0, \geq 2, \geq 0$ \\
signal b-jets                            & $\geq1$                    & =0                            & $\geq0, =0, \geq0$   \\
 $E^\mathrm{miss,rel.}_\mathrm{T}$       & $>100, 50, 40$             & 70--100                        & $>100, 50, 40$   \\
 other                                   &  -                         & -                             & -, $m_\textup{CT}$-veto, - \\
 \hline
\end{tabular}
\caption{\label{CR}{Requirements for entering each control region for top, $WW$ and $Z+X$ background estimation in the OS signal regions. These are used to estimate the top background in all OS signal regions, $WW$ in SR-OSjveto and $Z+X$ in all SF channels of the OS signal regions.
When each OS signal region requires differing control region definitions, the conditions are given as a comma separated list (SR-OSjveto, SR-2jets, SR-$m_\textup{T2}$). 
The $Z$-veto is a rejection of events with $m_{\ell\ell}$ within 10~GeV of the $Z$-mass (91.2~GeV), whereas the $Z$-window defines the reverse. In the $WW$ control region the $b$-jets considered are those with $p_\mathrm{T}>20$~GeV. The values quoted for $E^\mathrm{miss,rel}_\mathrm{T}$ are in units of~GeV.}}
\end{center}
\end{table}

\subsection{Backgrounds in SR-$m_\textup{T2}$}

In this letter, SR-$m_\textup{T2}$ is used to search for $\tilde{\ell}^\pm$ pair production and provides the best sensitivity to $\tilde{\chi}^\pm_1$ pair production. The main backgrounds in this region are: fully-leptonic $t\bar{t}$ and single top, $Z/\gamma^*+$jets and dibosons ($WW$, $WZ$ and $ZZ$). 

Fully-leptonic $t\bar{t}$ is comparable in size to the $WW$ background in all flavour channels. $Z/\gamma^*+$jets, $WZ$ and $ZZ$ processes (collectively, $Z+X$) are a small proportion of events in the DF channel, but comparable in size to the $WW$ and $t\bar{t}$ backgrounds in the SF channels. The remainder of the SM background is accounted for by fake lepton backgrounds. The methods used to evaluate these backgrounds in SR-$m_\textup{T2}$ are described in the following sections.

\subsubsection{Top}

The combined contribution from $t\bar{t}$ and single top events in each channel ($ee$, $e\mu$ or $\mu\mu$) is evaluated by normalising MC simulation to data in an appropriate control region.
Events in the control region (Table~\ref{CR}) must contain at least two signal jets, one of which must be $b$-tagged, and pass the requirement that $E^\mathrm{miss,rel.}_\mathrm{T}$ must be greater than 40~GeV.
The corresponding control region is dominated by top events. The contamination from non-top events is less than 4\%. The number of top events in the signal region ($N^\textup{SR}_\textup{top}$) is estimated from the number of data events in the control region ($N^\textup{CR}_\textup{top}$), after the subtraction of non-top backgrounds, using a transfer factor $\mathcal{T}$:
\begin{equation}
N^{\textup{SR}}_{\textup{X}}=N^{\textup{CR}}_{\textup{X}}\times\mathcal{T}\times \textup{S}_{\mathcal{T}}. \nonumber
\label{eq:CR_X_1}
\end{equation}
The factor, $\mathcal{T}$, the ratio of top events in the signal to those in the control regon is derived using MC.
\begin{equation}
\mathcal{T}=\left(\frac{N^\textup{SR}_\textup{X}}{N^\textup{CR}_\textup{X}}\right)_\textup{MC}.\nonumber
\label{eq:CR_X_2}
\end{equation}

The factor $\textup{S}_\mathcal{T}$ corrects for possible differences in jet-veto efficiency between data and MC simulation. Good agreement is observed in separate samples of $t\bar{t}$ and $Z/\gamma^*$+jets events and so this factor is taken to be equal to 1.0, with an uncertainty of 6\%.
The transfer factor is evaluated {before} the $m_\textup{T2}$ requirement is applied in the signal region since this requirement is designed to eliminate all but the tail of the $m_\textup{T2}$ distribution for $t\bar{t}$.
The efficiency of this requirement is then evaluated using MC simulation for a looser selection (which is assumed not to change the $m_\textup{T2}$ shape) and used to obtain the final estimate in SR-$m_\textup{T2}$. The efficiency of the $m_\textup{T2}$ requirement is found to be $\sim 2\%$ in each channel for top events with an uncertainty of $\sim50\%$. The uncertainty is largely dominated by MC statistical uncertainty, generator uncertainties and jet and lepton scales and resolutions. 

The evaluated $t\bar{t}$ components in each channel are consistent with pure MC estimates normalised to cross-sections to within 1$\sigma$. Data and MC simulation are also consistent at this level in the control region.
Negligible contamination from the SUSY signal models generated, in the region of the expected reach, is predicted.

\subsubsection{$Z+X$}

The $Z/\gamma^*+$jets background in the SF channels is also estimated by normalising MC simulation to data in a suitable control region. This procedure is important in order to handle appropriately 
possible detector imperfections affecting $E^\mathrm{miss}_\mathrm{T}$ measurement. This technique also estimates the $ZW$ and $ZZ$ components, providing a combined estimate of the total $Z+X$ background in the SF channels.

In the DF
channel the $Z/\gamma^*+$jets contribution is significantly smaller and arises mainly from $Z/\gamma^*\rightarrow\tau\tau$ 
decays. This and the diboson components of the $Z+X$ background in the DF channel are estimated using MC simulation. 

The control region (Table~\ref{CR}) used to estimate the $Z+X$ background in the SF channels is defined to be identical to the signal region but with the $Z$-veto reversed.
The normalisation is evaluated before the $m_\textup{T2}$ requirement, and the efficiency of the $m_\textup{T2}$ requirement is measured separately using MC simulation.
The population of data events inside the control region {not} produced by $Z+X$ processes is estimated using data $e\mu$ events inside the $Z$-window, correcting for the differences between electron and muon reconstruction efficiencies. This subtraction removes less than $2\%$ of the events in the control region.
This procedure also subtracts contributions from $Z/\gamma^*\rightarrow\tau\tau+$jets events which must be estimated using MC simulation.
The MC $m_\textup{T2}$ requirement efficiency for $Z+X$ events is taken to be 0.004 (0.003) for $e^+ e^-$ ($\mu^+\mu^-$) events with $\sim50\%$ uncertainty. 

The estimated $Z+X$ background is consistent within statistics with the MC prediction. No significant signal contamination is expected for the SUSY model points considered in the region of sensitivity for the searches reported in this letter.

\subsubsection{$WW$}
The $WW$ background is evaluated using MC normalised to cross-section and luminosity. The predictions from a variety of generators (see Section~\ref{sec:MC}) were compared before application of the $m_\textup{T2}$ requirement (to maximise acceptance for comparison), in order to assess the theoretical uncertainty on this estimate. The $m_\textup{T2}$ distribution in data agrees well with that in MC simulation, and the $E^\mathrm{miss,rel.}_\mathrm{T}$ region under consideration ($>40$~GeV) is close to the bulk of the $WW$ sample.

\subsubsection{Fake leptons}
In this letter the term ``fake leptons'' refers to both misidentified jets and real leptons that arise from decays or conversions.
The numbers of fake lepton events are estimated using the ``matrix method''~\cite{atlas-topxsec}.
First, fake leptons are identified as those satisfying a loose set of identification requirements corresponding to medium-level identification requirements and no isolation.
The real efficiency $r$ is calculated using data as the fraction of these loose leptons passing the signal lepton identification and isolation requirements in events with a lepton pair of mass lying within 5~GeV of the $Z$~boson~mass. 
The fake efficiency $f$ is calculated separately for misidentified jets or decays and conversions. 
The combined fake efficiency for misidentified jets or decays is calculated using MC events with $E^\mathrm{miss,rel.}_\mathrm{T}$ between 40 and 100~GeV, and validated using low-$E^\mathrm{miss,rel.}_\mathrm{T}$ regions in data. This region of moderate $E^\mathrm{miss,rel.}_\mathrm{T}$ is expected to give a sample composition that is representative of the various signal regions.
The fake efficiency for conversions is estimated in a data sample dominated by this process, with two muons of invariant mass within 10~GeV of 
the $Z$-mass, $E^\mathrm{miss,rel.}_\mathrm{T}<50$~GeV and at least one loose electron with $m_\textup{T}<40$~GeV (the conversion candidate).
The overall $f$ used is then the weighted (according to the relative proportions of each component present in the signal region) average of these two fake efficiencies.
Then, in the signal region the observed numbers of events in data with two loose leptons, two signal leptons, or one of each are counted. The number of events containing fake leptons in each signal region is finally obtained by acting on these observed counts with a $4\times 4$ matrix with terms containing $f$ and $r$ that relates real-real, real-fake, fake-real and fake-fake lepton event counts to tight-tight, tight-loose, loose-tight and loose-loose counts.

\subsection{Backgrounds in SR-OSjveto, SR-SSjveto and SR-2jets}

The same techniques are used to estimate the backgrounds in each remaining signal region, with two exceptions which are detailed in this section. Table~\ref{CR} details any changes to control region definitions used. 
\begin{enumerate}

\item Due to the high $E^\mathrm{miss,rel.}_\mathrm{T}$ requirement ($>100$~GeV) in SR-OSjveto,
  $WW$ is estimated using MC normalised to data in a control region. The control region used for its estimate is defined using the same requirements as the signal region but with slightly lower $E^\mathrm{miss,rel.}_\mathrm{T}$ (for orthogonality with the signal region) and an additional $b$-jet veto to suppress $t\bar{t}$ (Table~\ref{CR}). This control region is subject to a 24\% contamination from top events, which is estimated and removed using MC simulation.
  \item In SR-SSjveto, the leptons have the same charge, resulting in a generally different background composition, and the presence of an additional component: ``charge-flip''. The background components in this region are: fake leptons (estimated using the described matrix method), dibosons (estimated using MC events) and charge-flip. This mis-identification of charge arises when an electron in an event undergoes hard bremsstrahlung with subsequent photon conversion. The probability of an electron undergoing a flip is measured from $Z$ events in data using a likelihood technique~\cite{chargeflip}, and in MC simulation.
This probability, evaluated as a function of electron rapidity and $p_\mathrm{T}$, is applied to $t\bar{t}\rightarrow e^\pm {\ell}^\mp$, $Z+$jets and diboson MC events to evaluate the number of $e^\pm e^\pm$ and $e^\pm\mu^\pm$ events resulting from the charge-flip mechanism. The probability of misidentifying the charge of a muon is negligible. The possible double counting of charge-flip events in the matrix method for SR-SSjveto is not significant.
\end{enumerate}

\section{Systematic Uncertainties}

\label{sys}

In this analysis systematic uncertainties arise in the estimates of the background in 
the signal regions, as well as on the estimate of the SUSY signal itself.
The primary sources of systematic uncertainty are the jet energy scale (JES)~\cite{Aad:2011he} calibration, 
the jet energy resolution (JER)~\cite{jer}, choice of MC generator and lepton efficiencies and momentum measurements. Additional statistical uncertainties arise from limited numbers of MC and data events in the control and signal regions, and a 3.9\% luminosity uncertainty~\cite{lumi1,lumi2} for normalising MC events to cross-sections.

The JES has been
determined from a combination of test beam, simulation and in-situ measurements from 2011 $pp$ collision data.
Uncertainties on the lepton identification, momentum/energy
scale and resolution are estimated from samples of $Z\rightarrow l^+l^-$, $J/\psi\rightarrow l^+l^-$ and $W^\pm\rightarrow {\ell}^\mp\nu$ decays~\cite{muonefficiency,more_unc}. The
uncertainties on the jet and lepton energies are propagated to $E^\mathrm{miss,rel.}_\mathrm{T}$; an additional uncertainty on $E^\mathrm{miss}_\mathrm{T}$ arising from energy deposits not associated to any reconstructed objects is also included~\cite{ETmiss_unc}. 
Uncertainties on tfhe $b$-tagging efficiency are derived from data samples containing muons associated to jets~\cite{ATLAS-CONF-2011-102} using the method described in Ref.~\cite{system8}. Included are uncertainties in the mis-tag rate from charm~\cite{uncmistag}, and light flavour tagging~\cite{flavour}.

Theory and MC modelling uncertainties are evaluated for $t\bar{t}$ using the prescriptions described in Ref.~\cite{ISRFSR} (choice of generator, and ISR/FSR). For dibosons they are evaluated by varying the choice of generator. Theoretical uncertainties on the $Z/\gamma^*+$jets background from varying the PDF and renormalisation scales are also included.

When evaluating the fake lepton component in each region the dominant uncertainties arise from the dependency of the efficiencies on $E^\mathrm{miss,rel.}_\mathrm{T}$, differences between efficiencies obtained using OS and SS events and uncertainties in the relative normalisations of the different components. An additional uncertainty is applied based on differences observed in the fake efficiencies measured from data to validate the MC efficiencies if different validation regions are chosen.

The relative sizes of these sources of systematic uncertainty are detailed in Table~\ref{sys_summary}. In SR-$m_\textup{T2}$, the jet and lepton energy scales and resolutions are the most significant uncertainties.
In SR-OSjveto and SR-2jets, where $t\bar{t}$ and $WW$ are the most significant SM backgrounds (accounting for approximately 80--85\% of the SM contribution), the uncertainties in the MC modelling dominate. In SR-SSjveto, because of the significant fake component, the error on the fake estimate from the sources described becomes the only significant source of uncertainty. 

\begin{table}[!thp]
\small
\begin{center}
\begin{tabular}{|l|c|c|c|c|}
\hline
SR- & $m_\textup{T2}$ & OSjveto & SSjveto & 2jets  \\
\hline
Total statistical & 9 & 4 &13 & 6  \\
Total systematic & 19 & 19 &  35 &  49\\
\hline
Jet uncertainties & 9 & 8  & 3 & 5\\
Lepton uncertainties & 14 & 1 & 1 & 5 \\
$b$-tagging efficiency & 1 & 1 & 0 & 14\\
MC modelling & 7 & 17 & 4 & 45 \\
Fake leptons &  5 & 5 &  35 & 4 \\
\hline
\end{tabular}
\end{center}
\caption{\label{sys_summary}Systematic uncertainties (\%) on the total background estimated in each signal region for all flavours combined. 
The total statistical uncertainty includes limited MC event numbers in the control and signal regions. Jet systematic uncertainties include: JES, JER and $E^\mathrm{miss}_\mathrm{T}$ cluster and pile-up uncertainties. Lepton systematic uncertainties include: all lepton scales and resolutions, reconstruction and trigger efficiencies. MC modelling uncertainties include choice of generator, ISR/FSR and modelling of the $Z/\gamma^*+$jets line-shape.}
\end{table}

In the SUSY mass planes, the theoretical uncertainty on each of the signal cross-sections is included.
These arise from considering the  cross-section envelope defined using the 68\% CL ranges of the {\tt CTEQ6.6} and {\tt MSTW 2008 NLO} PDF sets, and independent variations of the factorisation and renormalisation scales (see Section~\ref{sec:MC}).
Further uncertainties on the numbers of predicted signal events arise from the various experimental uncertainties.

\section{Results and Interpretation}

\begin{figure*}[!tp]
\begin{center}
\subfigure[]{\includegraphics[scale=0.38]{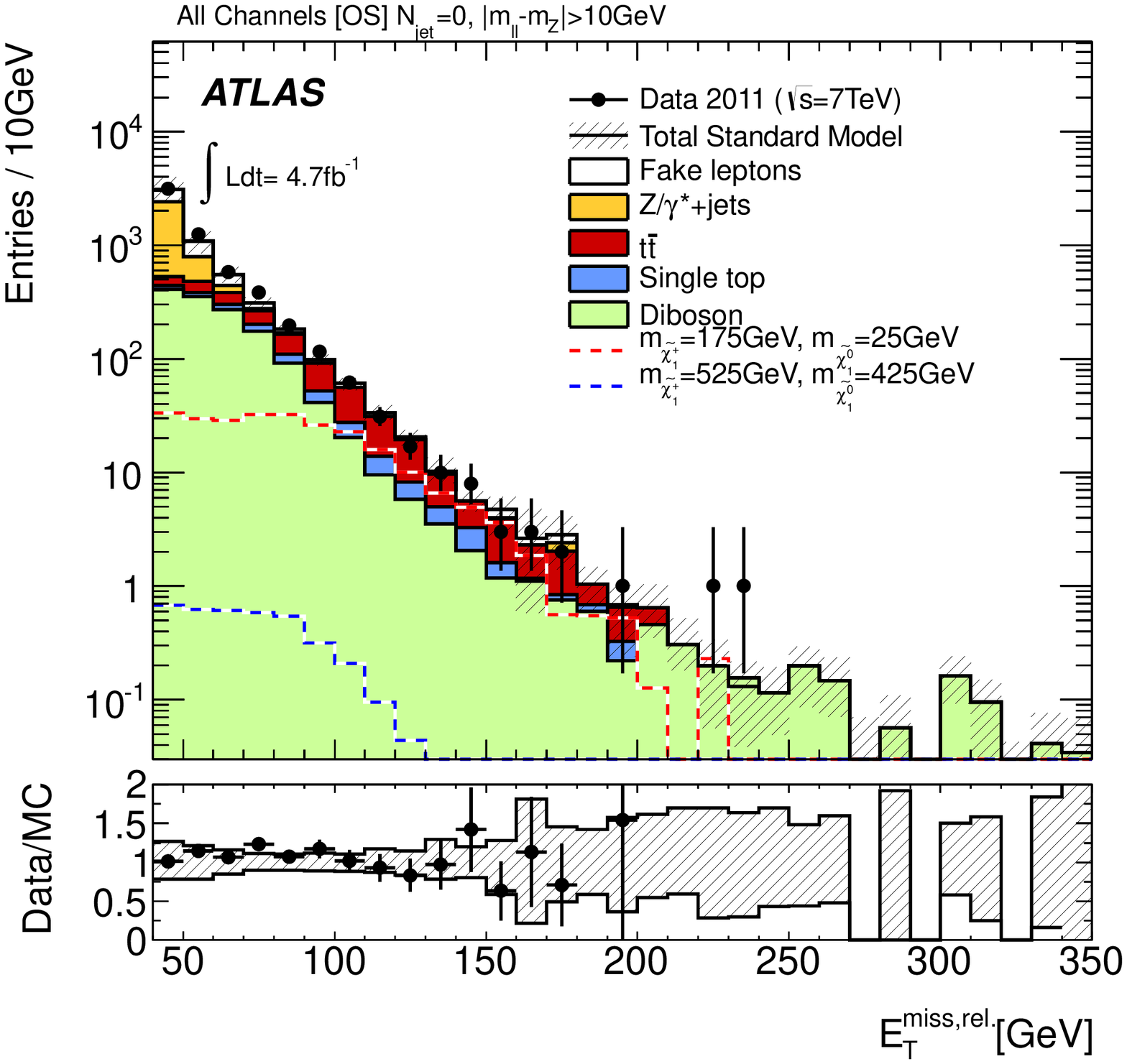}}
\subfigure[]{\includegraphics[scale=0.38]{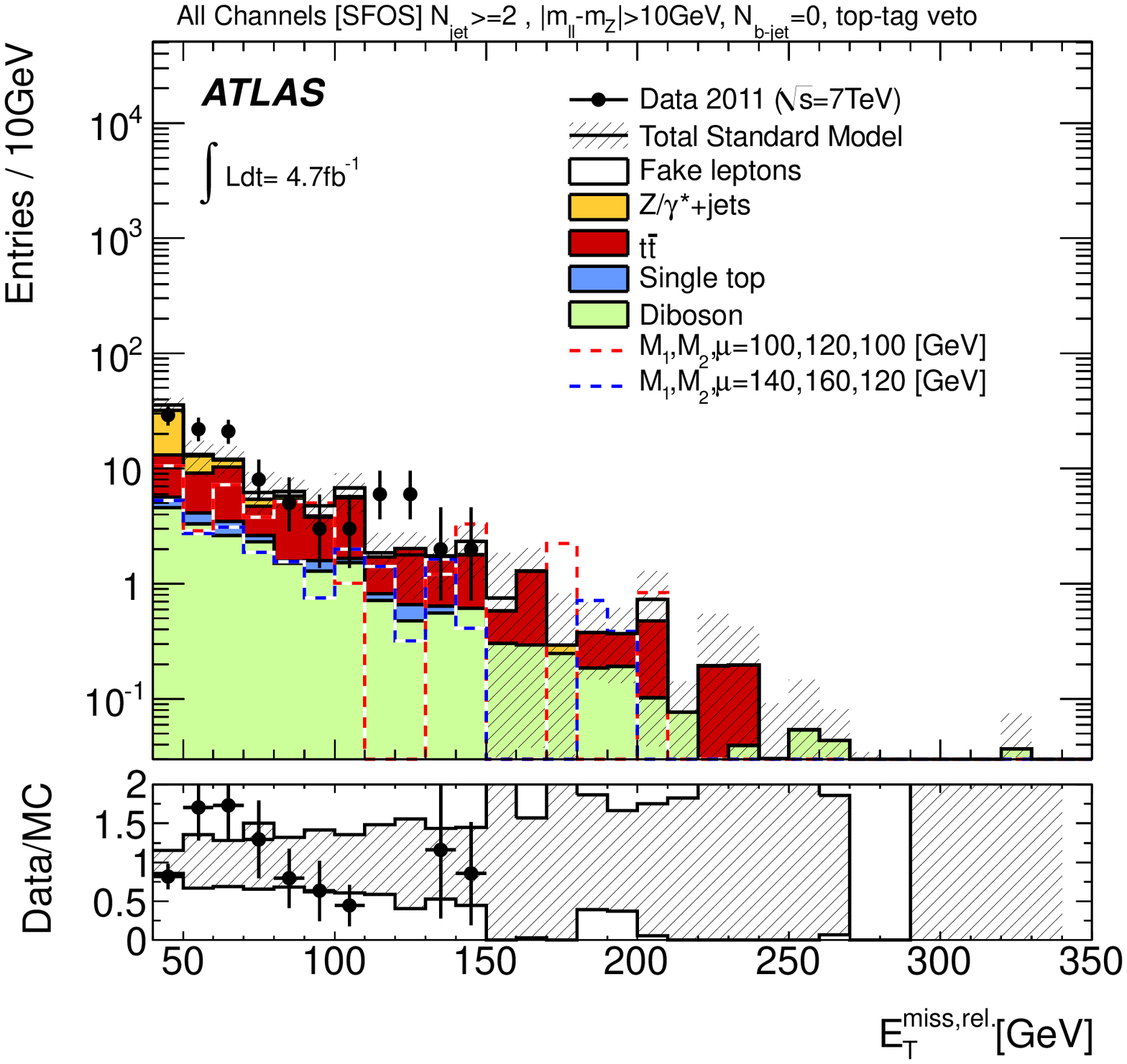}}
\subfigure[]{\includegraphics[scale=0.38]{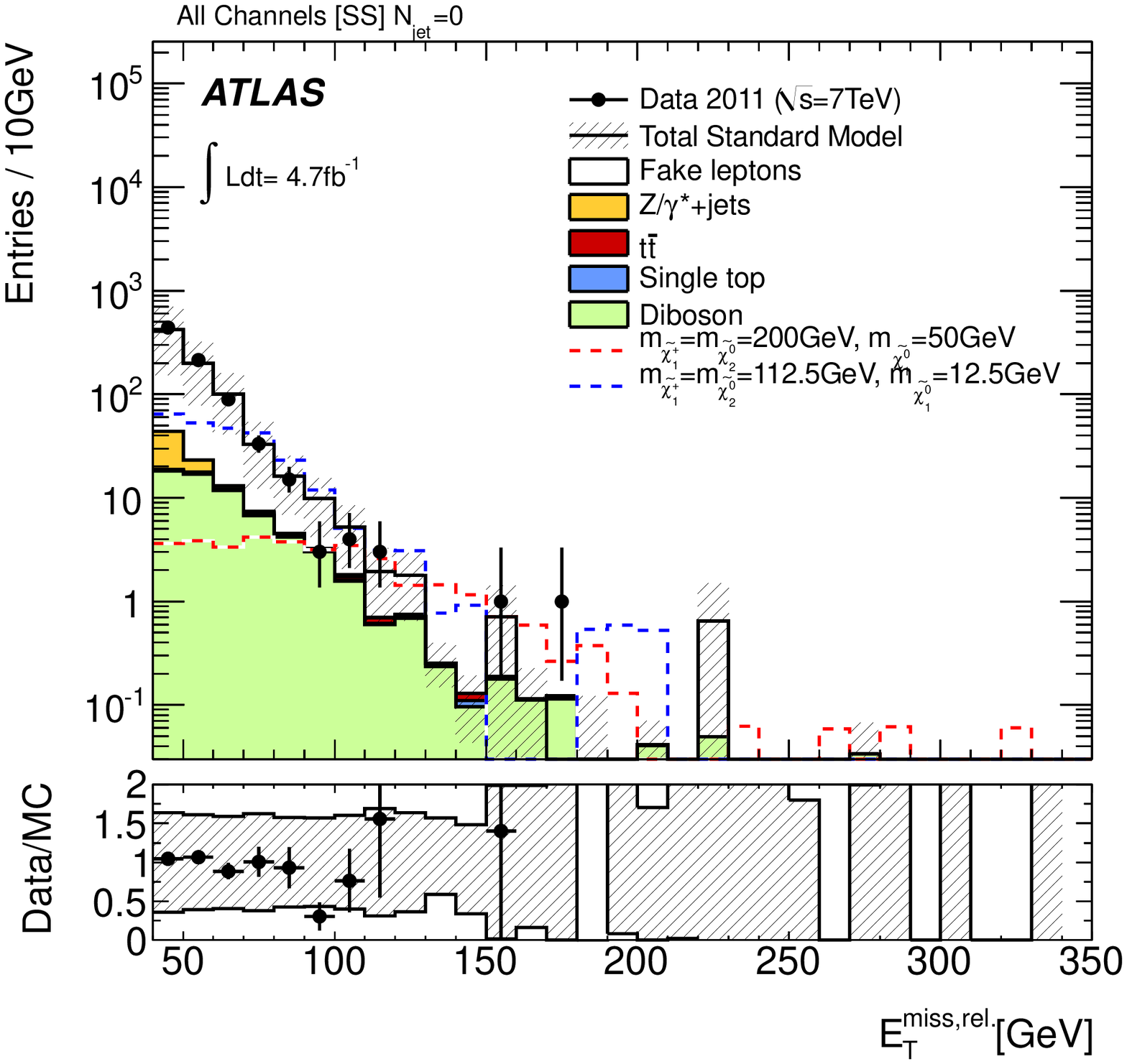}}
\subfigure[]{\includegraphics[scale=0.38]{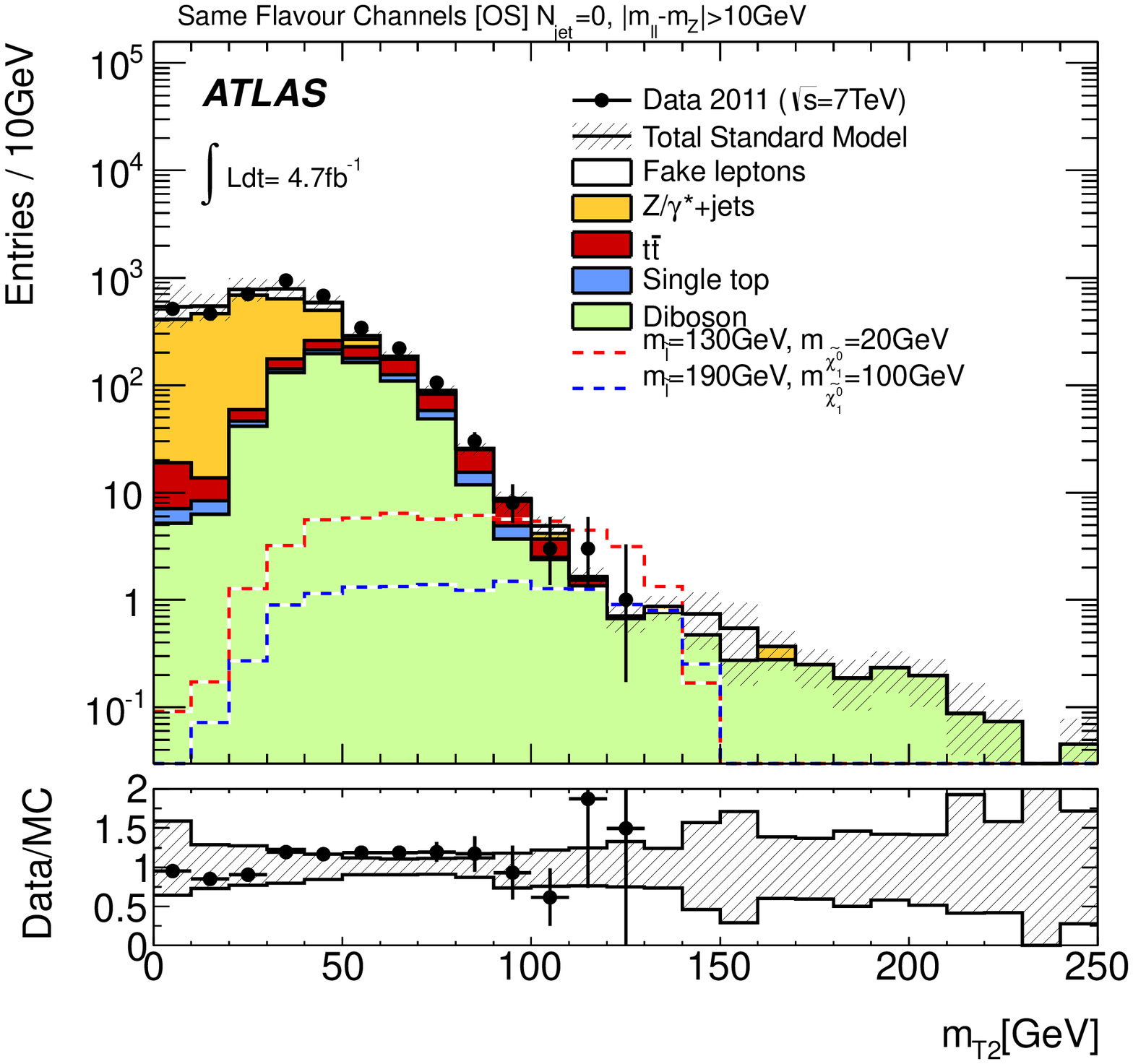}}
\caption{The $E^\mathrm{miss,rel.}_\mathrm{T}$ distributions prior to the final requirement on $E^\mathrm{miss,rel.}_\mathrm{T}$ for (a)~SR-OSjveto, (b)~SR-2jets and (c)~SR-SSjveto, and (d)~$m_\textup{T2}$ in SR-m$_\textup{T2}$, prior to the application of the $m_\textup{T2}$ requirement. In~(d) only the SF channels are shown. The hatched bands indicate the experimental uncertainties on the background expectations. All components are from MC except for that labelled ``Fake leptons''. The contribution labelled ``Diboson'' accounts for $WW$, $WZ$ and $ZZ$ processes.
The bottom panels show the ratio of the data to the expected background (points) and the systematic uncertainty on the background (shaded area). In each figure two signal points are illustrated. In (d) two models of direct slepton pair production are illustrated, with ($\tilde{\ell},\tilde{\chi^0_1}$) masses of (130,20) and (190,100)~GeV. In (a) the two points illustrated are for $\tilde{\chi}^\pm_1\tilde{\chi}^\mp_1$ production in the simplified model with ($\tilde{\chi}^\pm_1$,$\tilde{\chi}^0_1$) masses of (175,25) and (525,425)~GeV. In (c) the simplified model points illustrated have ($\tilde{\chi}^\pm_1$,$\tilde{\chi}^0_1$) masses of (200,50) and (112.5,12.5)~GeV. In (b) two pMSSM model points with masses ($M_1,M_2,M_\mu$) of (100,120,100) and (140,160,300)~GeV are illustrated.
\label{leppt_distributions}
}
\end{center}
\end{figure*}

Fig.~\ref{leppt_distributions} illustrates the level of agreement in each signal region, prior to the application of the final requirement on $E^\mathrm{miss,rel.}_\mathrm{T}$ and $m_\textup{T2}$, between the data and the SM prediction. 
For each signal region two illustrative model points are also presented.

Table~\ref{results} compares the observations in data in each flavour channel and in each signal region with the evaluated background contributions. Good agreement is observed across all channels and in each signal region. The absence of evidence for SUSY weak production allows limits to be set on the visible cross-section for non-SM physics in each signal region, $\sigma_{\rm vis}=\sigma \times \varepsilon \times A$, for which this analysis has acceptance $A$ and efficiency $\varepsilon$. These are calculated using the modified frequentist CL$_\textup{s}$ prescription~\cite{Read:2002hq} by comparing the number of observed events in data with the SM expectation using the profile likelihood ratio as test statistic.
All systematic uncertainties and their correlations are taken into account via nuisance parameters.

\begin{table*}[!hbp]
\small
\begin{center}
\begin{tabular}{|c|c|c|c|c|c|}
 \hline 
 \multicolumn{6}{|c|}{SR-$m_\textup{T2}$}\\  
 \hline
         & $e^+ e^-$ & $e^\pm \mu^\mp$ & $\mu^+\mu^-$ & all & SF \\
 \hline
 $Z+$X                                   & $3.2\pm1.1\pm1.7$ & $0.3\pm0.1\pm0.2$  & $3.6\pm1.3\pm1.7$     & $7.1\pm1.7\pm2.1$  & $6.8\pm1.7\pm2.1$ \\
 $WW$                                    & $2.3\pm0.3\pm0.4$ & $4.8\pm0.4\pm0.7$  & $3.5\pm0.3\pm0.5$     & $10.6\pm0.6\pm1.5$ & $5.8\pm0.4\pm0.9$\\
 $t\bar{t}$, single top                  & $2.6\pm1.2\pm1.3$ & $6.2\pm1.6\pm2.9$  & $4.1\pm1.3\pm1.6$     & $12.9\pm2.4\pm4.6$ &$6.8\pm1.8\pm2.3$ \\
 Fake leptons                            & $1.0\pm0.6\pm0.6$ & $1.1\pm0.6\pm0.8$  & $-0.02\pm0.01\pm0.05$ & $2.2\pm0.9\pm1.4$  & $1.0\pm0.6\pm0.6$\\
 \hline
 Total                                   & $9.2\pm1.8\pm2.5$ & $12.4\pm1.7\pm3.1$ & $11.2\pm1.9\pm3.0$    & $32.8\pm3.2\pm6.3$ & $20.4\pm2.6\pm3.9$\\
 Data                                    & 7                 & 9                  & 8                     & 24                 & 15 \\
 $\mathrm{\sigma^{obs(exp)}_{vis}}$ (fb) & 1.5 (1.8)         & 1.6 (2.0)          & 1.6 (1.9)             & 2.5 (3.3)          & 1.9 (2.5)\\
 \hline
 \hline
 \multicolumn{6}{|c|}{SR-OSjveto}\\           
 \hline
                                         & $e^+ e^-$          & $e^\pm \mu^\mp$     & $\mu^+\mu^-$       & \multicolumn{2}{|c|}{all} \\
 \hline
 $Z+$X                                   & $4.5\pm1.2\pm1.2$  & $3.0\pm0.9\pm0.5$   & $4.7\pm1.1\pm1.2$  & \multicolumn{2}{|c|}{$12.2\pm1.8\pm1.8$} \\
 $WW$                                    & $8.8\pm1.8\pm4.4$  & $20.9\pm2.6\pm6.2$  & $13.3\pm1.9\pm3.5$ & \multicolumn{2}{|c|}{$43.0\pm3.7\pm12.2$} \\
 $t\bar{t}$, single top                  & $21.1\pm2.3\pm4.2$ & $47.7\pm3.4\pm20.5$ & $27.5\pm2.5\pm9.0$ & \multicolumn{2}{|c|}{$96.2\pm4.8\pm29.5$}\\
 Fake leptons                            & $2.9\pm1.2\pm1.2$  & $6.9\pm1.8\pm2.6$   & $0.4\pm0.6\pm0.3$  & \multicolumn{2}{|c|}{$10.3\pm2.2\pm4.1$}\\
 \hline
 Total                                   & $37.2\pm3.3\pm6.4$ & $78.5\pm4.7\pm20.9$ & $45.9\pm3.4\pm9.4$ & \multicolumn{2}{|c|}{$161.7\pm6.7\pm30.8$}  \\
 Data                                    & 33                 & 66                  & 40                 & \multicolumn{2}{|c|}{139} \\
 $\mathrm{\sigma^{obs(exp)}_{vis}}$ (fb) & 3.3 (3.8)          & 6.8 (7.8)           & 4.0 (4.6)          & \multicolumn{2}{|c|}{9.8 (11.9)} \\
 \hline
   \hline
  \multicolumn{6}{|c|}{SR-SSjveto}\\       
 \hline
                                         & $e^{\pm} e^{\pm}$    & $e^\pm \mu^\pm$      & $\mu^{\pm}\mu^{\pm}$ &\multicolumn{2}{|c|}{all} \\
 \hline
 Charge flip                             & $0.49\pm0.03\pm0.17$ & $0.34\pm0.02\pm0.11$ & ---                  & \multicolumn{2}{|c|}{$0.83\pm0.04\pm0.18$} \\
 Dibosons                                & $0.62\pm0.13\pm0.18$ & $1.93\pm0.23\pm0.36$ & $0.94\pm0.16\pm0.26$ & \multicolumn{2}{|c|}{$3.50\pm0.31\pm0.54$} \\
 Fake leptons                            & $3.2\pm0.9\pm1.7$    & $2.9\pm0.9\pm1.9$    & $0.6\pm0.6\pm0.3$    & \multicolumn{2}{|c|}{$6.6\pm1.4\pm3.8$}\\
 \hline 
 Total                                   & $4.3\pm0.9\pm1.7$    & $5.1\pm1.0\pm1.9$    & $1.5\pm0.6\pm0.4$    & \multicolumn{2}{|c|}{$11.0\pm1.5\pm3.9$}\\
 Data                                    & 1                    & 5                    & 3                    & \multicolumn{2}{|c|}{9}\\
 $\mathrm{\sigma^{obs(exp)}_{vis}}$ (fb) & 0.7 (1.1)            & 1.6 (1.6)            & 1.3 (0.9)            & \multicolumn{2}{|c|}{1.9 (2.1)}\\
 \hline
 \hline
  \multicolumn{6}{|c|}{SR-2jets}\\            
 \hline
                                         & $e^+ e^-$           & $e^\pm \mu^\mp$ & $\mu^+\mu^-$        & \multicolumn{2}{|c|}{SF} \\
 \hline 
 $Z+$X                                   & $3.8\pm1.3\pm2.7$   & ---             & $5.8\pm1.6\pm3.9$   & \multicolumn{2}{|c|}{$9.6\pm2.0\pm5.1$}\\
 $WW$                                    & $6.4\pm0.5\pm4.3$   & ---             & $8.4\pm0.6\pm5.7$   & \multicolumn{2}{|c|}{$14.8\pm0.7\pm9.9$}\\
 $t\bar{t}$, single top                  & $14.8\pm1.9\pm9.2$  & ---             & $22.1\pm2.1\pm20.7$ & \multicolumn{2}{|c|}{$36.9\pm2.9\pm29.6$}\\
 Fake leptons                            & $2.5\pm1.2\pm1.5$   & ---             & $1.7\pm1.3\pm0.8$   & \multicolumn{2}{|c|}{$4.2\pm1.8\pm2.3$}\\
 \hline
 Total                                   & $27.5\pm2.6\pm10.6$ & ---             & $37.9\pm3.0\pm21.0$ & \multicolumn{2}{|c|}{$65.5\pm4.0\pm31.8$}\\
 Data                                    & 39                  & ---             & 39                  & \multicolumn{2}{|c|}{78}\\
 $\mathrm{\sigma^{obs(exp)}_{vis}}$ (fb) & 6.9 (5.3)           & ---             & 7.7 (7.6)           & \multicolumn{2}{|c|}{13.6 (12.5)} \\
 \hline
 \hline
\end{tabular}\caption{\label{results}Evaluated SM backgrounds in each signal region separated by flavour ($ee$, $e\mu$, $\mu\mu$) and combined in an ``all'' channel. In SR-$m_\textup{T2}$ the evaluated background components in the SF channel are quoted separately as the $e\mu$ channel is not appropriate for a direct slepton search. The second quoted error is the total systematic uncertainty whereas the first is the statistical uncertainty arising from limited numbers of MC events. The effect of limited data events in the control region is included in the systematic uncertainty. In all OS signal regions and channels the component $Z+$X includes the contributions from $Z/\gamma^*+$jets, $WZ$ and $ZZ$ events. All statistical uncertainties are added in quadrature whereas the systematic uncertainties are obtained after taking full account of all correlations between sources, backgrounds and channels. Quoted also are the observed (expected) 95\% confidence limits on the visible cross-section for non-SM events in each signal region, $\mathrm{\sigma^{obs(exp)}_{vis}}$.}\end{center}
\end{table*}

\begin{figure*}
\subfigure[]{\includegraphics[width=0.5\textwidth]{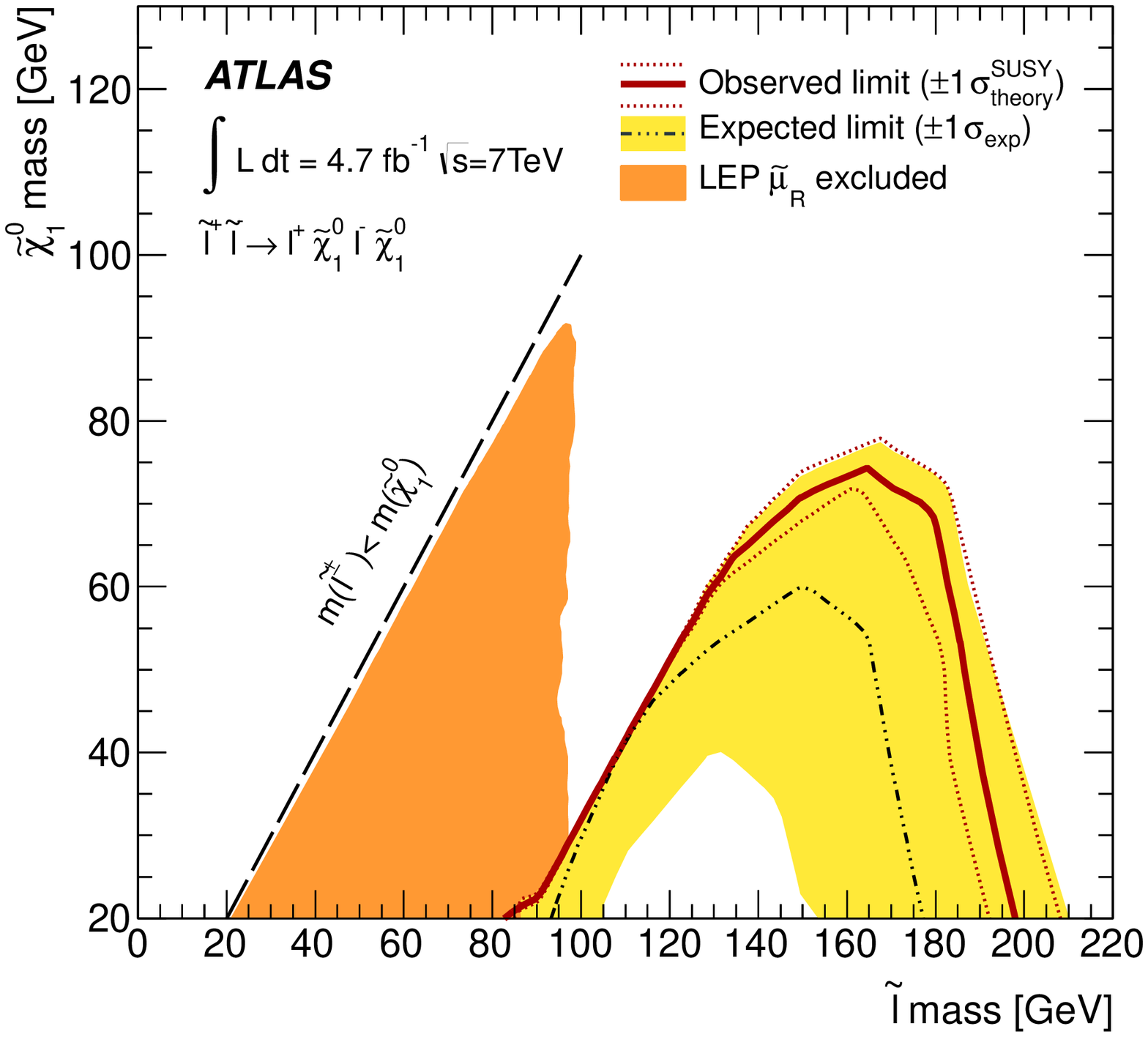}}
\subfigure[]{\includegraphics[width=0.5\textwidth]{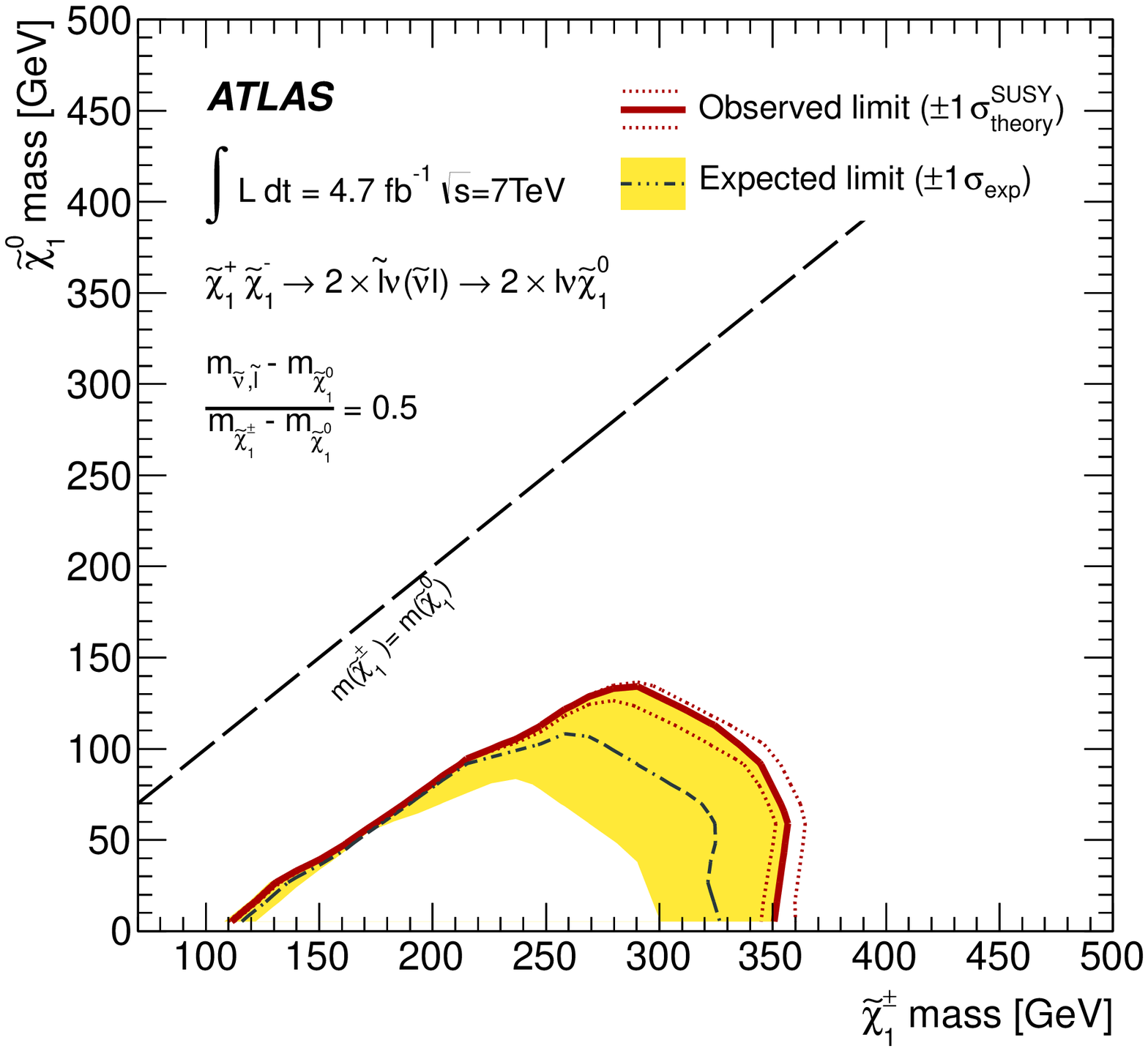}}
\caption{\label{interpretation_slepton}95\% CL exclusion limits for $\tilde{\ell}^\pm$ pair production in the 
$m_{\tilde{\ell}}-m_{\tilde{\chi}^0_1}$ mass plane of (a) the direct slepton pMSSM and (b) $\tilde{\chi}^\pm_1\tilde{\chi}^\mp_1$ 
pair production in the simplified model. The dashed and solid lines show the 95\% CL$_\textup{s}$ 
expected and observed limits, respectively, including all uncertainties except for the theoretical signal 
cross-section uncertainty (PDF and scale). The solid band around the expected limit shows the $\pm1\sigma$ result where 
all uncertainties, except those on the signal cross-sections, are considered. The $\pm1 \sigma$ lines around the observed 
limit represent the results obtained when moving the nominal signal cross-section up or down by the $\pm 1 \sigma$ 
theoretical uncertainty. Illustrated also in~(a) is the LEP limit~\cite{lepsusy} on the mass of the right-handed 
smuon, $\tilde{\mu}_{R}$. The LEP limit is a conservative limit on slepton pair production: if right-handed slepton masses are excluded, left-handed sleptons of equivalent masses are automatically excluded.}
\end{figure*}

The direct slepton pair production 95\% CL exclusion region is shown 
in Fig.~\ref{interpretation_slepton}(a) in the neutralino--slepton mass plane, using 
the results of SR-$m_\textup{T2}$ in the SF channel. Shown are the 95\% CL$_\textup{s}$ 
expected (dashed black) and observed limits (solid red) obtained by including all
uncertainties except the theoretical signal cross-section uncertainty. The solid yellow band indicates the impact 
of the experimental uncertainties on the expected limits whereas the dashed red lines around the observed 
limit show the changes in the observed limit as the nominal signal cross-sections are scaled up and 
down by the 1$\sigma$ theoretical uncertainties. A common value for left-handed electron and left-handed 
smuon mass between 85 and 195~GeV is excluded when the lightest neutralino has a mass of 20~GeV. 
The sensitivity decreases as the value of $m_{\tilde{\ell}}-m_{\tilde{\chi}^0_1}$ decreases and gives rise 
to end-points in the $m_\textup{T2}$ distribution at lower mass, nearer to the end-points of the SM backgrounds. 
For a 60~GeV neutralino only sleptons with masses between 135 and 180~GeV are excluded. 

The direct $\tilde{\chi}^\pm_1$ pair production limits are set for the simplified model,  
in the scenario of wino-like charginos decaying into the lightest neutralino via an intermediate 
on-shell charged slepton. The best expected limits are obtained by using for each signal point the signal region 
that provides the {best expected} $p$-value. The resulting limit for
$\tilde{\chi}^\pm_1\tilde{\chi}^\mp_1$ production is illustrated in Fig.~\ref{interpretation_slepton}(b).
Chargino masses between 110 and 340 GeV are excluded at 95\% CL for a 10~GeV neutralino.
The best sensitivity is provided by SR-$m_\textup{T2}$. Previous gaugino searches at the Tevatron and
the LHC~\cite{cdftrilepton,D0trilepton,2lep2011,3lep2011} focused on $\tilde{\chi}^\pm_1 \tilde{\chi}^0_2$
associated production. The present result provides a new mass limit on $\tilde{\chi}^\pm_1$  independently
of the mass of the $\tilde{\chi}^0_2$. 

\begin{figure*}[!tp]
\subfigure[~$M_1=100$~GeV]{\includegraphics[width=0.5\textwidth]{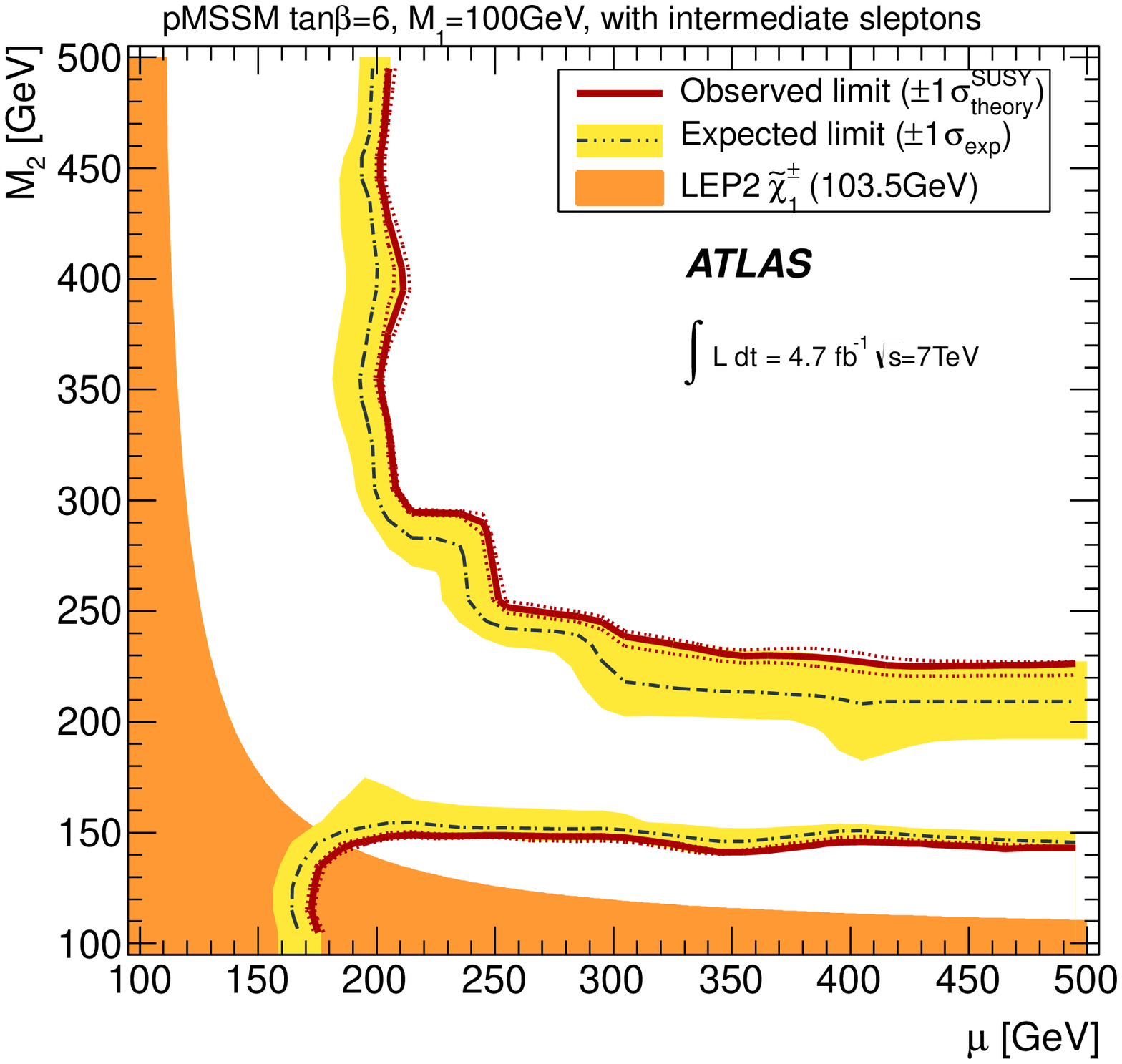}}
\subfigure[~$M_1=140$~GeV]{\includegraphics[width=0.5\textwidth]{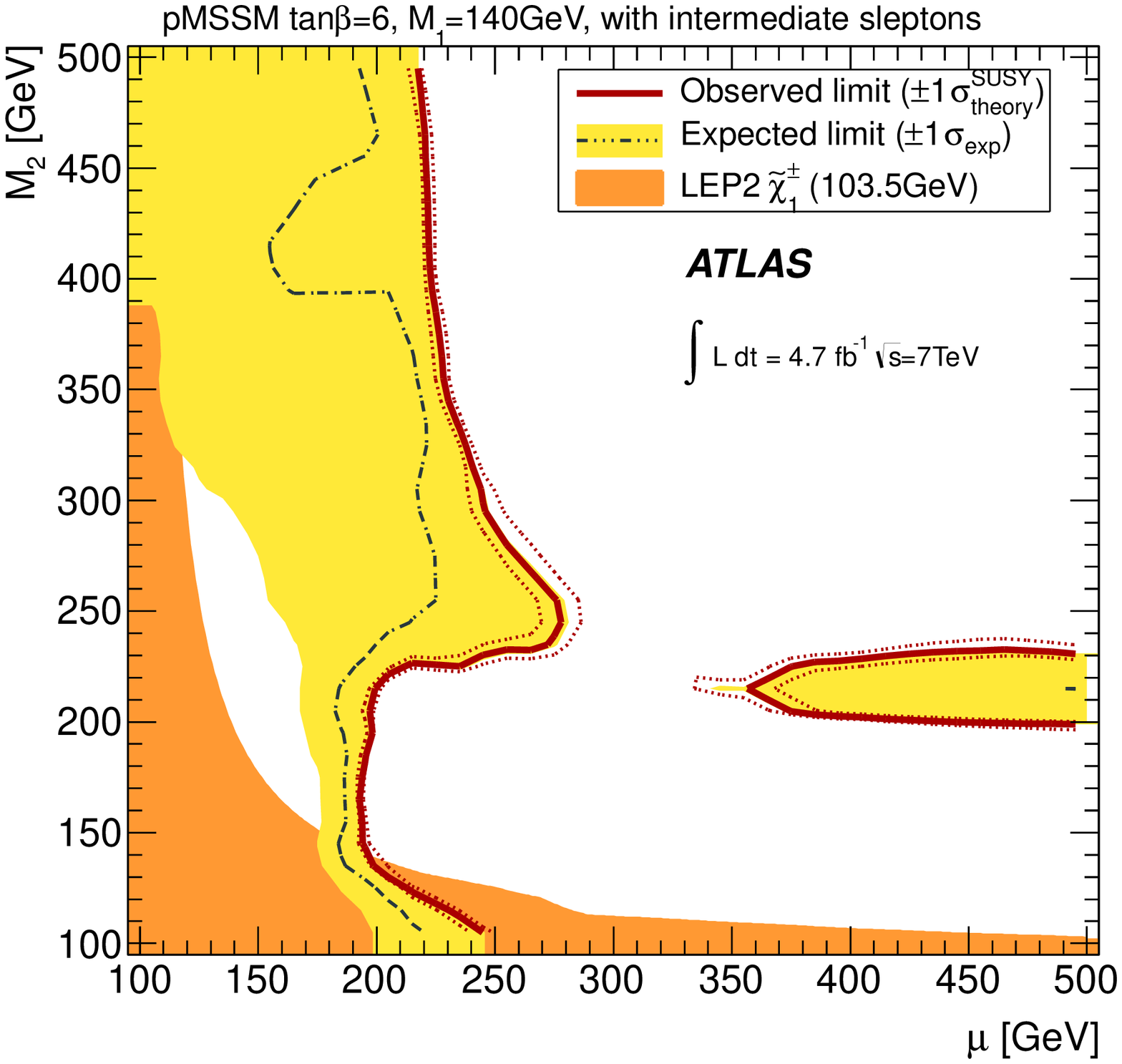}}
\begin{center}
\subfigure[~$M_1=250$~GeV]{\includegraphics[width=0.5\textwidth]{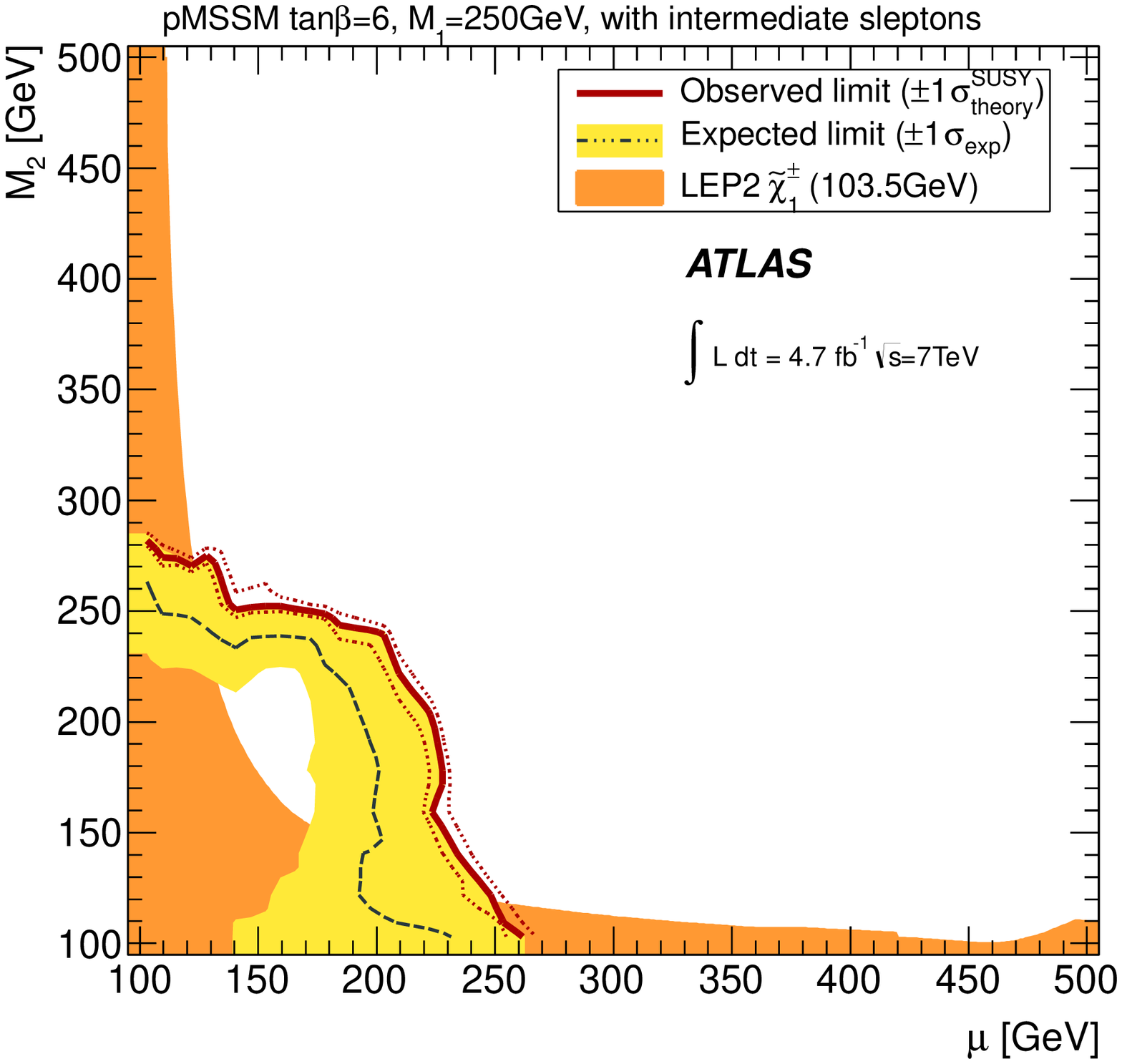}}
\end{center}
\caption{\label{interpretation_pmssm}95\% CL exclusion limits in the $\mu$--$M_2$ mass plane of the pMSSM for 
(a) $M_1=100$~GeV, (b) $M_1=140$~GeV and (c) $M_1=250$~GeV. The dashed and solid lines show the 
95\% CL$_\textup{s}$ expected and observed limits, respectively, including all uncertainties except 
for the theoretical signal cross-section uncertainty (PDF and scale). The solid band around the expected limit 
shows the $\pm1\sigma$ result where all uncertainties, except those on the signal cross-sections, are considered. 
The $\pm1 \sigma$ lines around the observed limit represent the results obtained when moving the nominal signal 
cross-section up or down by the $\pm 1 \sigma$ theoretical uncertainty.}
\end{figure*}

The signal regions are combined in Fig.~\ref{interpretation_pmssm} to derive exclusion limits in the pMSSM 
$\mu-M_{2}$ plane for $\tan{\beta}=6$,  by selecting for each signal point the signal region which provides the {best expected} $p$-value.
Figs.~\ref{interpretation_pmssm}(a)--(c) show respectively the exclusion limits for $M_{1}=100,~140,~250~$~GeV. 
The present result significantly extends previous limits in the pMSSM $\mu-M_{2}$ plane. The model independent limits in 
Table~\ref{results} provide additional constraints on other gaugino production 
channels discussed previously in this letter. In particular, SR-2jets provides sensitivity to models 
where one gaugino produced in association with $\tilde{\chi}^{0}_{2}$ decays hadronically. The best sensitivity 
to models where final states containing $\geq3$ leptons dominate would come from a statistical combination of the results set 
in SR-2jets, SR-OSjveto and SR-SSjveto, and results of searches for three or more leptons~\cite{3lep2011}.

\section{Summary}

This letter has presented a dedicated search for $\tilde{\ell}^\pm$ and $\tilde{\chi}^\pm_1$ pair production in final states 
with two leptons and $E^\mathrm{miss}_\mathrm{T}$. In scenarios where sleptons decay directly into the lightest neutralino 
and a charged lepton, left-handed slepton masses between 85 and 195~GeV for a 20~GeV neutralino are excluded at 95\% confidence.
In the scenario of chargino pair production, with wino-like charginos decaying into the lightest neutralino via an intermediate 
on-shell charged slepton, chargino masses between 110 and 340~GeV are excluded at 95\% CL for a neutralino of 10~GeV. 
New limits in the pMSSM $\mu-M_{2}$ plane are provided for $\tan{\beta}=6$. Signal regions targeting several gaugino production 
and decay modes into two-lepton final states have also been used to set limits on the visible cross-section. 
\section{Acknowledgements}

We thank CERN for the very successful operation of the LHC, as well as the
support staff from our institutions without whom ATLAS could not be
operated efficiently.

We acknowledge the support of ANPCyT, Argentina; YerPhI, Armenia; ARC,
Australia; BMWF, Austria; ANAS, Azerbaijan; SSTC, Belarus; CNPq and FAPESP,
Brazil; NSERC, NRC and CFI, Canada; CERN; CONICYT, Chile; CAS, MOST and NSFC,
China; COLCIENCIAS, Colombia; MSMT CR, MPO CR and VSC CR, Czech Republic;
DNRF, DNSRC and Lundbeck Foundation, Denmark; EPLANET and ERC, European Union;
IN2P3-CNRS, CEA-DSM/IRFU, France; GNSF, Georgia; BMBF, DFG, HGF, MPG and AvH
Foundation, Germany; GSRT, Greece; ISF, MINERVA, GIF, DIP and Benoziyo Center,
Israel; INFN, Italy; MEXT and JSPS, Japan; CNRST, Morocco; FOM and NWO,
Netherlands; RCN, Norway; MNiSW, Poland; GRICES and FCT, Portugal; MERYS
(MECTS), Romania; MES of Russia and ROSATOM, Russian Federation; JINR; MSTD,
Serbia; MSSR, Slovakia; ARRS and MVZT, Slovenia; DST/NRF, South Africa;
MICINN, Spain; SRC and Wallenberg Foundation, Sweden; SER, SNSF and Cantons of
Bern and Geneva, Switzerland; NSC, Taiwan; TAEK, Turkey; STFC, the Royal
Society and Leverhulme Trust, United Kingdom; DOE and NSF, United States of
America.

The crucial computing support from all WLCG partners is acknowledged
gratefully, in particular from CERN and the ATLAS Tier-1 facilities at
TRIUMF (Canada), NDGF (Denmark, Norway, Sweden), CC-IN2P3 (France),
KIT/GridKA (Germany), INFN-CNAF (Italy), NL-T1 (Netherlands), PIC (Spain),
ASGC (Taiwan), RAL (UK) and BNL (USA) and in the Tier-2 facilities
worldwide.

\bibliographystyle{atlasnote}
\bibliography{DGSlepton2Lep-journal}

\clearpage
\begin{flushleft}
{\Large The ATLAS Collaboration}

\bigskip

G.~Aad$^{\rm 48}$,
T.~Abajyan$^{\rm 21}$,
B.~Abbott$^{\rm 111}$,
J.~Abdallah$^{\rm 12}$,
S.~Abdel~Khalek$^{\rm 115}$,
A.A.~Abdelalim$^{\rm 49}$,
O.~Abdinov$^{\rm 11}$,
R.~Aben$^{\rm 105}$,
B.~Abi$^{\rm 112}$,
M.~Abolins$^{\rm 88}$
% ...
\bigskip
\\
% ...
$^{11}$ Institute of Physics, Azerbaijan Academy of Sciences, Baku, Azerbaijan\\
$^{12}$ Institut de F{\'\i}sica d'Altes Energies and Departament de F{\'\i}sica de la Universitat Aut{\`o}noma de Barcelona and ICREA, Barcelona, Spain\\
% ...
$^{21}$ Physikalisches Institut, University of Bonn, Bonn, Germany\\
% ...
$^{48}$ Fakult{\"a}t f{\"u}r Mathematik und Physik, Albert-Ludwigs-Universit{\"a}t, Freiburg, Germany\\
$^{49}$ Section de Physique, Universit{\'e} de Gen{\`e}ve, Geneva, Switzerland\\
% ...
$^{88}$ Department of Physics and Astronomy, Michigan State University, East Lansing MI, United States of America\\
% ...
$^{105}$ Nikhef National Institute for Subatomic Physics and University of Amsterdam, Amsterdam, Netherlands\\
% ...
$^{111}$ Homer L. Dodge Department of Physics and Astronomy, University of Oklahoma, Norman OK, United States of America\\
$^{112}$ Department of Physics, Oklahoma State University, Stillwater OK, United States of America\\
% ...
$^{115}$ LAL, Universit{\'e} Paris-Sud and CNRS/IN2P3, Orsay, France\\
% ...
\end{flushleft}


\end{document}